# Sonochemical synthesis of large two-dimensional $Bi_2O_2CO_3$ nanosheets for hydrogen evolution in photocatalytic water splitting

Tushar Gupta,[1] Nicole Rosza,[1] Markus Sauer,[2] Alexander Goetz,[3] Maximilian Winzely,[4] Jakob Rath,[1] Shaghayegh Naghdi,[1] Dogukan H. Apaydin,[1] Gernot Friedbacher,[4] Annette Foelske,[2] Sarah M. Skoff,[3] Bernhard C. Bayer,[1,*] Dominik Eder[1,*]

[1]*Institute of Materials Chemistry, Technische Universität Wien (TU Wien), Getreidemarkt 9/BC, Vienna, A-1060, Austria*

[2]*Analytical Instrumentation Center, Technische Universität Wien (TU Wien), Lehargasse 6/BL, Vienna, A-1060, Austria*

[3]*Atominstitut, Technische Universität Wien (TU Wien), Stadionallee 2, Vienna, A-1020, Austria*

[4]*Institute of Chemical Technologies and Analytics, Technische Universität Wien (TU Wien), Getreidemarkt 9/BB, Vienna, A-1060, Austria*

*\*Corresponding authors: bernhard.bayer-skoff@tuwien.ac.at (B. C. Bayer), dominik.eder@tuwien.ac.at (D. Eder)*

**Abstract**

Laterally large (~3 µm), atomically-thin two-dimensional (2D) $Bi_2O_2CO_3$ nanosheets (2D bismuth oxycarbonate, 2D bismutite) are fabricated via sonochemically-assisted template-free synthesis. Key to the synthesis of the freestanding, laterally large 2D $Bi_2O_2CO_3$ nanosheets from bulk Bi powder is choice of suspension medium, controlled reaction temperatures and several hours processing time. Lateral sizes of 2D $Bi_2O_2CO_3$ can be controlled between µm-sized nanosheets and tens of nm sized nanoflakes solely based on the choice of suspension medium. The here introduced 2D $Bi_2O_2CO_3$ nanosheets/-flakes are then hybridized by a simple mix-and-match approach with $TiO_2$ nanoparticles for testing in suspension-type photocatalytic hydrogen production via water splitting. This introduces the 2D $Bi_2O_2CO_3$ with $TiO_2$ as a promising noble-metal-free co-catalyst for photocatalytic hydrogen evolution. Our results enrich the fabrication toolbox of emerging 2D pnictogen oxycarbonates towards large 2D nanosheets and demonstrate the promising potential of 2D $Bi_2O_2CO_3$ as an advantageous (co-)catalyst for hydrogen evolution in photocatalytic water splitting.

**Introduction**

Two-dimensional (2D) pnictogens such as 2D antimony ("antimonene") and 2D bismuth ("bismuthene") have spurred tremendous interest in electronics, energy applications and catalysis.[1–4] Likewise, more recently, also 2D pnictogen compounds have come into focus because of their combination of unusual, potentially useful properties in electronics, energy and catalysis.[5–12] For the case of bismuth, this includes recent work on binary 2D bismuth oxides[9–12] and 2D ternary and multinary oxygen-containing bismuth compound phases, incl. 2D Bi-oxyhalides, 2D $Bi_2WO_6$, 2D $Bi_2MoO_6$ or 2D $BiVO_4$.[2,3,8,13,14]

Amongst the ternary bismuth compound phases, the bismuth oxycarbonate (BOC) $Bi_2O_2CO_3$ phase, also called bismutite and bismuth subcarbonate $(BiO)_2CO_3$, is of particular interest.[15] $Bi_2O_2CO_3$ has an intrinsically layered structure composed of alternating $[Bi_2O_2]^{2+}$ and $[CO_3]^{2-}$ sub-layers and is a semiconductor with a band gap of ~3.1 – 3.5 eV.[15] In nanostructured form $Bi_2O_2CO_3$ has been shown to have useful properties in particular towards energy, catalysis and photocatalysis.[15] In particular for photocatalysis, 2D morphology can offer benefits over other morphologies incl. intrinsically high specific surface areas and short migration lengths of photogenerated charge carriers to the reaction fronts on the 2D materials' surfaces. This can reduce recombination losses and thus lead to higher activity of 2D forms of common materials.[16] Synthesis of $Bi_2O_2CO_3$ in atomically-thin 2D morphology remains however underdeveloped, in particular in terms of rational control over lateral sizes and thicknesses.[15,17–48]

Towards filling this gap, we here report a simple method of sonochemically-assisted[49,50] template-free synthesis of laterally large, atomically-thin 2D $Bi_2O_2CO_3$ nanosheets. In particular, our synthesis approach readily allows control over lateral 2D $Bi_2O_2CO_3$ size between unusually large, µm-sized nanosheets and small, tens of nm sized nanoflakes solely depending on suspension medium.

While nanostructured $Bi_2O_2CO_3$ is a popular photocatalyst for organic pollutant and NO degradation,[15,17–33,46–48] it has to date received little attention for photocatalytic solar fuel production such as in photocatalytic hydrogen evolution reaction (HER) from water splitting.[34,35,51] This is despite photocatalytic HER being one of the key technologies to meet the demands of the growing energy crisis in a sustainable way.[52,53] Photocatalytic HER critically hinges on the availability of (cost-)efficient, scalable (co-)catalysts and the search for such (co-)catalysts is an ongoing challenge.[52,53]

To this end, we here hybridize our newly introduced 2D $Bi_2O_2CO_3$ nanosheets/-flakes by a simple mix-and-match approach with $TiO_2$ nanoparticles and test these hybrids as heterogeneous photocatalysts in suspension-type photocatalytic HER from water splitting. Our results thereby introduce 2D $Bi_2O_2CO_3$ with $TiO_2$ as a promising noble-metal-free (co-)catalyst for photocatalytic HER.

## Results and Discussion

**Synthesis of 2D $Bi_2O_2CO_3$ nanosheets/-flakes.** Our sonochemically-assisted synthesis (Supporting Fig. S1) first encompasses liquid-phase-exfoliation (LPE)[49,50] type treatment of bulk Bi powder which has a layered, buckled rhombohedral β-Bi/A7 crystal structure[54] and a particle size of up to 150 µm (Supporting Fig. S2). To provide energy to the system we employ ultrasonic bath immersion.[49,50] We undertake our sonication with air-tight closed sample vials that are immersed in the temperature-controlled sonication bath. Importantly, we screen 5 different suspension media ("solvents") for the Bi during our sonochemical treatment, namely water ($H_2O$), methanol (MeOH), isopropanol (IPA), ethanol (EtOH) and (motivated by prior literature[55]) IPA:$H_2O$ (4:1) mixture. All solvents are of technical grade and are used without particular purging to remove dissolved gases before synthesis. A wide screening of solvents is often key in LPE and related techniques as 2D materials' suspendabilities are known to be highly solvent dependent.[49] Important for our recipe development, ultrasonication employs very long processing time (15 h) and very close control of bath temperature during sonication to either "room temperature" (RT, 22 °C to 26 °C) or "ice bath" conditions (ICE, 0.1 °C to 2.2 °C). Subsequent to sonication, suspensions are centrifuged and then supernatant is collected for further characterization and photocatalytic testing. We note that after the sonication step the suspension are visibly opaque compared to parent solvents, but that after the centrifugation step the suspensions look virtually as translucent as their parent solvents to the naked eye (Supporting Fig. S3). This suggests that concentrations of material in centrifuged suspensions is low. Gravimetric quantification of concentration of solid content via solvent evaporation corroborates this and put concentrations in our suspension to a maximum of ~50 mg/l (which is the detection limit of our balance system). A detailed description of our synthesis procedure is given in the Supporting Information.

Fig. 1 shows scanning electron microscopy (SEM) images of obtained products from sonication and centrifugation (after drop casting onto $SiO_2$/Si wafer) for MeOH (Fig. 1a,b) and $H_2O$ (Fig. 1d,e) from RT and ICE processing. The SEM images confirm that, despite the visually translucent suspensions, ample solid material is suspended in our sonicated and centrifuged samples. Strikingly, while MeOH (both in ICE and RT) produced small nanoflakes (lateral sizes ~80nm), $H_2O$ resulted in large 2D nanosheets of roughly square shapes with lateral sizes of several µm (ICE: average 3.5 µm ± 2.6 µm standard deviation; RT: average 2.3±1.3 µm) and an apparently low thickness (as flakes appear electron transparent in the SEM images). IPA (RT, ICE), EtOH (RT, ICE) and IPA: $H_2O$ (RT) lead to small nanoflakes (Supplementary Fig. S4a-e) akin to MeOH, which for IPA and EtOH also appear to be thicker compared to $H_2O$ and MeOH. IPA:$H_2O$ (ICE) produces thin square nanosheets with µm lateral sizes (Supplementary Fig. S4f).

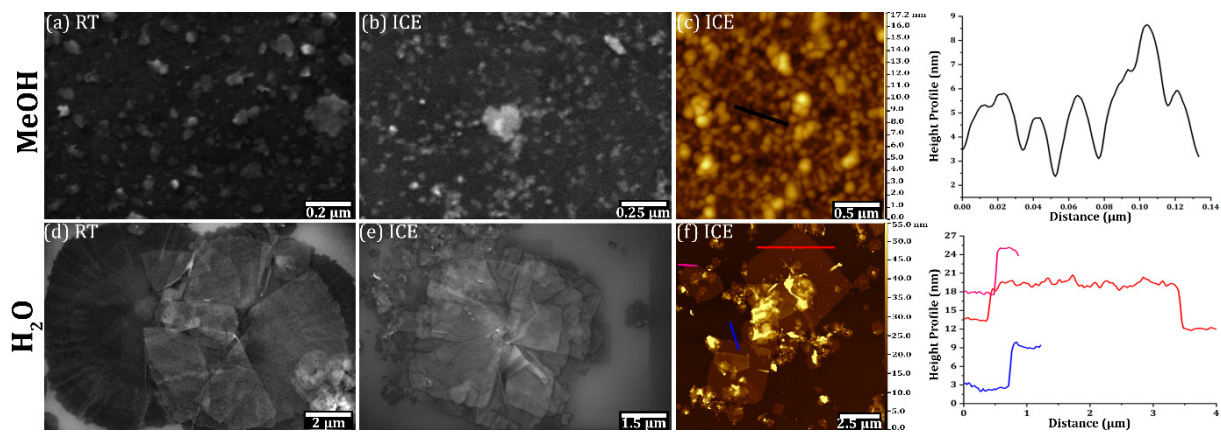

**Figure 1.** (a,b) SEM micrographs of MeOH RT and ICE, (c) AFM image of MeOH ICE. (d,e) SEM images of H$_2$O RT and ICE (f) AFM image of H$_2$O ICE. Line profiles over the flakes are plotted alongside the AFM images in (c,f) to obtain thickness measurements of the flakes.

In Fig. 1c,f we present atomic force microscopy (AFM) measurements to ascertain the thickness of the laterally small MeOH ICE and laterally large H$_2$O ICE products. Fig. 1c confirms that MeOH ICE results in thin (~2.5 – 3 nm) but laterally small (only ~80nm) nanoflakes. In very stark contrast, H$_2$O ICE in Fig. 1f resulted in approximately square-shaped nanosheets with ultra-high aspect ratios, with lateral sizes of ~3 µm and atomic-scale thicknesses of only 4 – 6 nm. Supporting Figure S5 shows that MeOH RT results in correspondingly small nanoflake morphology and that H$_2$O RT results in correspondingly large, atomically thin 2D nanosheet morphology.

Fig. 2a-c presents a structural characterization of the MeOH ICE products by bright-field (BF) transmission electron microscopy (TEM, overview Fig. 2a, lattice resolution side-view in Fig. 2b and lattice resolution top-view in Fig. 2c with Fourier Transforms (FTs) in insets). TEM confirms the small lateral size, atomic thickness of the nanoflakes and additionally proves their crystalline nature. Fig. 2d-f shows corresponding TEM data for the H$_2$O ICE products in terms of BF-TEM overview (Fig. 2d), lattice resolution TEM in top-view (Fig. 2e, FT in inset) and a selected area electron diffraction pattern (SAED) from top-view (Fig. 2f). For the H$_2$O ICE preparations, TEM confirms the large lateral size, atomic thickness and approximate square shape, and additionally proves the single crystalline nature of the large nanosheets.

To identify the crystal structure of the nanoflakes and nanosheets from the lattice resolution TEM and SAED data, we consider that the starting educt is Bi, and that reactions to Bi$_x$O$_y$ and Bi$_x$O$_y$C$_z$ via oxygen and carbon from MeOH and H$_2$O decomposition in sonication and via dissolved oxygen and carbonate ions CO$_3^{2-}$ from dissolved CO$_2$ in the solvents are likely reaction pathways during our long sonication processes. Interestingly, comparing against multiple Bi, Bi$_x$O$_y$ and Bi$_x$O$_y$C$_z$ database structures (see extended discussion in Supporting Information), we consistently find that the best match to our experimental top-view TEM/SAED data (Fig. 2c,e,f) is orthorhombic Bi$_2$O$_2$CO$_3$ viewed along the [010] zone axis (Fig. 2g, left). (Please note that instead of the orthorhombic unit cell alternatively often a tetragonal unit cell is used to describe Bi$_2$O$_2$CO$_3$. We discuss our results using this tetragonal notation in the Supporting Information.) Our nanoflakes and nanosheets are thus layered orthorhombic Bi$_2$O$_2$CO$_3$ with the alternating Bi$_2$O$_2$ and CO$_3$ sub-layers parallel to the substrate (Fig. 2g, middle, orthorhombic $Bi_2O_2CO_3$ with [010] or (010) texture). Beyond structure identification from top-view images, for the MeOH ICE we also obtained lattice resolved side-view images of the 2D flakes (Fig. 2b). These confirm their atomically-thin "few-layer" nature and we measure lattice fringes in the side view of ~0.68 nm, which is also in excellent agreement of the layer distance in orthorhombic $Bi_2O_2CO_3$ of ~0.68 nm (corresponding to (040) plane family, see also side-view SAED/FT simulation in Supporting Fig. S6). Thus our side-view TEM data also corroborates 2D $Bi_2O_2CO_3$ with (010) texture. Our TEM-based structural phase assignment is also backed up on a larger scale by Bragg-Brentano X-ray diffractometry (XRD) measurements of nanosheets drop-cast onto $SiO_2$/Si wafers (Supporting Fig. S7). The measured XRD patterns are highly consistent with $Bi_2O_2CO_3$ with (040) (and equivalently (010)) texture. Notably, XRD is inconsistent with other possible candidate phases, most importantly excluding all $Bi_2O_3$ polymorphs (see Supporting Information for extended discussion). Furthermore, XRD does not indicate the presence of any other phases, thus confirming phase purity of the 2D $Bi_2O_2CO_3$ nanosheets from our synthesis procedure. The phase purity is also corroborated by the number of flakes/sheets observed at lattice resolution in TEM (7 for MeOH, 15 for $H_2O$) which are all consistently best indexed to 2D $Bi_2O_2CO_3$ along [010] zone axis. Similarly, we find that MeOH RT and $H_2O$ RT synthesis products are best matched by 2D $Bi_2O_2CO_3$ (Supporting Fig. S8). Our phase assignment therefore suggests that during 15 h sonication, the Bi has not only transformed to 2D nanoflake/-sheet morphology[50] but concurrently got oxidized and bridged/intercalated by $CO_3^{2-}$ to form 2D $Bi_2O_2CO_3$. We will discuss a suggested mechanism for this process below.

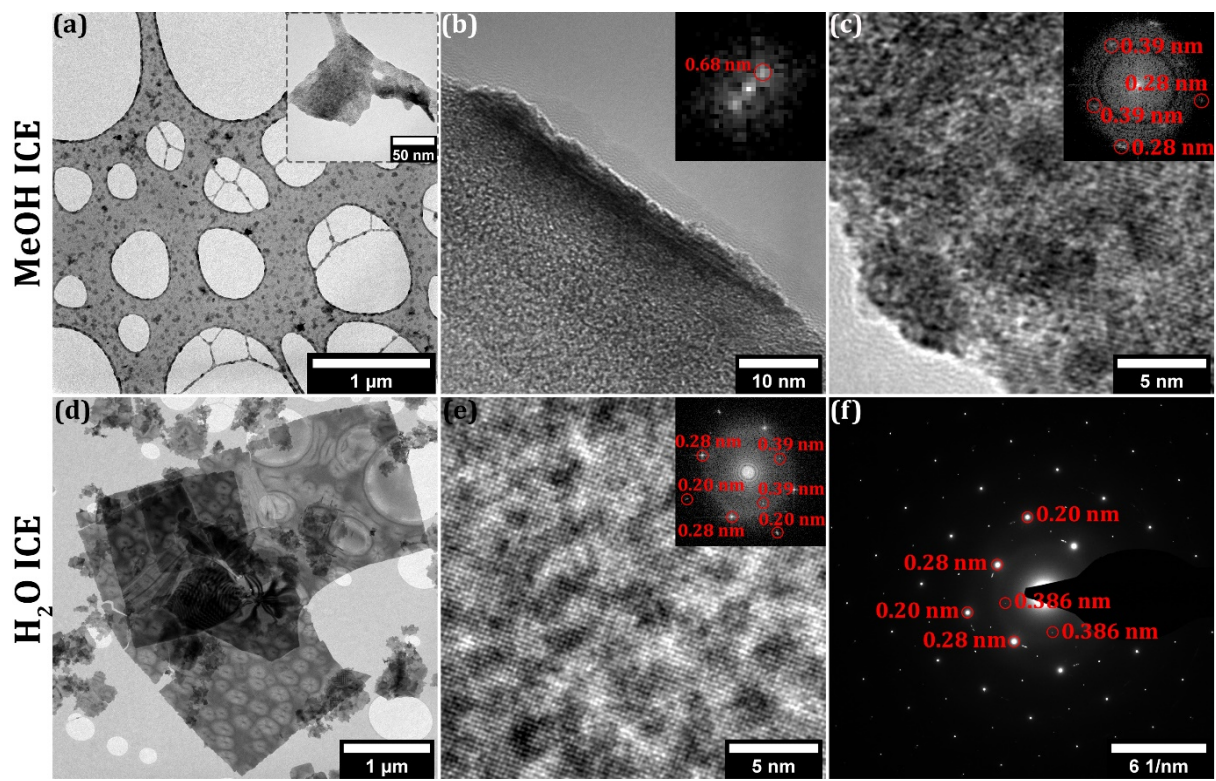

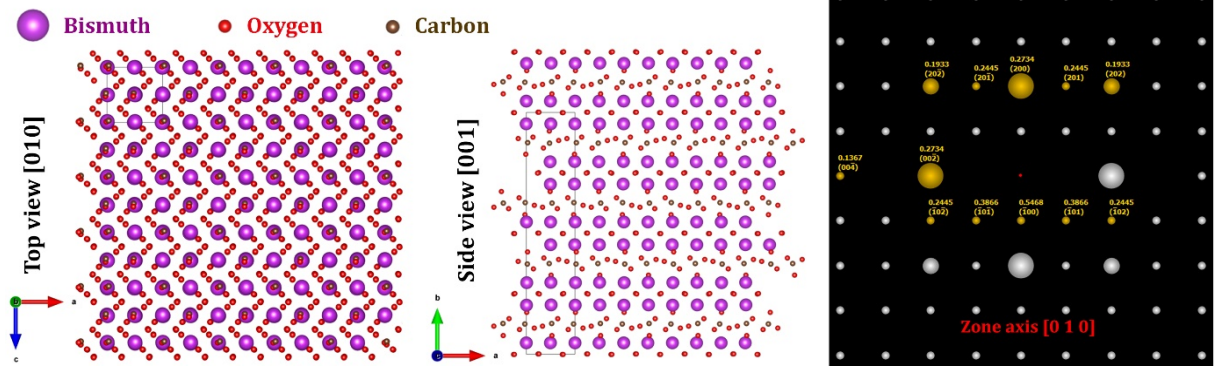

**Figure 2.** (a-c) MeOH ICE studied by overview BF-TEM (a), lattice resolution side-view (b, FT in inset) and lattice resolution top-view (c, FT inset) BF-TEM of nanoflake. (d-f) H$_2$O ICE studied by overview (d) and lattice resolution (e, FT in inset, top-view of flake) BF-TEM and top-view SAED (f). (g) shows a structural model of orthorhombic 2D Bi$_2$O$_2$CO$_3$ (powder diffraction file: 04-009-8533)[56] in top-view along [010] (left) and side-view along [001] (middle) and a simulated SAED/FT pattern (middle) for Bi$_2$O$_2$CO$_3$ along [010] zone axis (i.e. top-view) with salient reflections matches to experimental data in (e,f) highlighted. For a corresponding SAED/FT simulation of side-view see Supporting Fig. S6.

Additional characterization data of our 2D Bi$_2$O$_2$CO$_3$ nanoflakes/-sheets is presented in Supporting Figures S9 to S13. Energy dispersive X-ray spectroscopy (EDX) mapping in SEM (Supporting Fig. S9) clearly shows the homogeneous presence of Bi, O and C in H$_2$O ICE nanosheets, fully consistent with the 2D Bi$_2$O$_2$CO$_3$ assignment. X-ray photoelectron spectroscopy (XPS) measurements of the H$_2$O ICE samples (Supporting Fig. S10-S12) confirm Bi4f, O1s and C1s signatures[18] consistent with Bi$_2$O$_2$CO$_3$ and exclude the presence of metallic Bi.[50] Photoluminescence (PL) measurements of $H_2O$ (ICE) nanosheets and MeOH (ICE) flakes on $SiO_2$/Si in Supporting Fig. S13 reveal a PL emission peak at ~550 nm (for 532 nm excitation), consistent with prior reports of PL signatures of $Bi_2O_2CO_3$.[29,57] Combined, TEM/SAED, XRD, EDX, XPS and PL all corroborate the assignment of our $H_2O$ nanosheets and MeOH nanoflakes to 2D $Bi_2O_2CO_3$.

We note however that we observe a strong solvent dependence not only in terms of resulting 2D $Bi_2O_2CO_3$ nanoflake/-sheet morphology as between MeOH vs. $H_2O$, but also in terms of resulting phase when comparing with IPA, EtOH and IPA:$H_2O$ solvents. While for MeOH and $H_2O$ we obtain 2D $Bi_2O_2CO_3$ nanoflakes/-sheets, as described above, for IPA (RT, ICE) amorphous nanoflakes are obtained and for EtOH amorphous (ICE) or β-Bi(001) (RT) nanoflakes are obtained (Supporting Fig. S14). In turn for IPA:$H_2O$ (RT) again 2D $Bi_2O_2CO_3$ nanosheets are obtained, while for IPA:$H_2O$ (ICE) amorphous nanoflakes result (Supporting Fig. S15). This underlines an active role of the solvent in the transformation, oxidation and carbonate bridging/intercalation of initial Bi into the 2D $Bi_2O_2CO_3$ nanoflakes/-sheets, with best results towards crystalline 2D $Bi_2O_2CO_3$ in $H_2O$ for large nanosheets and MeOH for small nanoflakes, respectively.

Comparing to prior literature, the observation of 2D $Bi_2O_2CO_3$ is surprising: Recent work by Pumera et al.[12,50] found for a similar sonication scheme to ours and also in water, not 2D $Bi_2O_2CO_3$ but large metallic β-Bi[42-1] nanosheets to be produced while MeOH resulted in small metallic β-Bi(001) nanoflakes. In fact, Pumera et al. explicitely excluded bismuth oxidation in their study. Key differences were however the excitation energy source (tip sonication[12,50] vs. bath sonication here) and in particular a much shorter sonication time (max. 60 min[12,50] vs. 15 h here). Pumera et al. suggested for their 60 min sonications result in $H_2O$ not only physical exfoliation of the parent Bi powder to take place but due to Bi's low melting point (~271 °C) Bi melting under the tip sonicator, dissolution effects and then crystallite growth in suspension to partake in large nanosheet formation.[12,50] In contrast, in MeOH for 60 min Pumera et al. suggested fragmentation (rather than crystallite growth as in $H_2O$) to be the dominant mechanism for small nanoflake formation. While Pumera et al.'s max. 60 min processing times resulted in metallic β-Bi nanosheets/-flakes, we suggest that for our much longer 15 h processing in addition to the crystallite growth/fragmentation mechanisms suggested by Pumera et al. also concurrent oxidation and carbonate bridging/intercalaction processes of the Bi in $H_2O$ and MeOH take place. Compared to many prior reports on $Bi_2O_2CO_3$ synthesis,[15] we notably do not employ a dedicated source of oxygen or carbonate ions in our synthesis but only Bi as dedicated precursor in the various solvents. We thus suggest that the sources of the required oxygen and carbon for our proposed mechanism are coming from solvent decomposition under ultrasound and dissolved atmospheric oxygen and carbonate $CO_3^{2-}$ ions from dissolved atmospheric $CO_2$ in the solvents (see also Supporting Information).[36–38,40,58] Thus bismuth oxycarbonate nanosheets and -flakes are produced under our conditions in $H_2O$ and MeOH, respectively. This oxidation and carbonate bridging/intercalation process was arrested in the Pumera et al. work due to their shorter processing time.

In prior reports on 2D $Bi_2O_2CO_3$, lateral sizes of atomically-thin 2D $Bi_2O_2CO_3$ were largely limited at ~1 μm,[15,17–48] while our work increases lateral nanosheet sizes to ~3 μm. Our report so far establishes a controllable synthesis route for laterally large (~3 μm), atomically-thin 2D $Bi_2O_2CO_3$ nanosheets ($H_2O$ ICE and RT) and for laterally small (~ 80 nm), atomically-thin 2D $Bi_2O_2CO_3$ nanoflakes (MeOH ICE and RT), importantly solely based on the choice of solvent.

**Photocatalytic testing of 2D $Bi_2O_2CO_3$ nanoflakes/nanosheets.** Since, nanostructured $Bi_2O_2CO_3$ has been prior reported to have interesting properties as photocatalyst,[15] in the remainder of this report we investigate the here introduced 2D $Bi_2O_2CO_3$ nanosheets and nanoflakes as heterogeneous (co-)catalysts in suspension-type photocatalytic HER from water splitting under ultra-violet (UV) excitation. To date, $Bi_2O_2CO_3$ has predominantly been screened as a photocatalyst for organic pollutant degradation and NO removal.[15,17–33,46–48] To this end, not only monolithic $Bi_2O_2CO_3$ alone but also various mixed-dimensionality hybrids of $Bi_2O_2CO_3$ with metals, metal-oxides, sulfides, Bi-compounds and carbon-nitride (g-$C_3N_4$) have been investigated.[15,21–23,25,26,35,43,51,59–61] Only very little work has however thus far focused on photocatalytic HER from $Bi_2O_2CO_3$,[15] namely only hybrids consisting of $Bi_2O_2CO_3$ nanoplates/Pt,[34] $Bi_2O_2CO_3$ nanoparticles/g-$C_3N_4$/Pt[51] and $Bi_2O_2CO_3$ nanoplates/Bi.[35] Further, despite the archetypical importance of $TiO_2$ as photocatalyst,[52,53] only very few reports on potentially synergetic performance from $Bi_2O_2CO_3$/$TiO_2$ hybrids exist.[15,59] Few prior examples are $Bi_2O_2CO_3$ nanoflowers hybridized with $TiO_2$ nanoparticles on graphene sheets[59] and $Bi_2O_2CO_3$ nanoplates hybrids with $TiO_2$ nanoparticles and carbon networks,[60] whereby both hybrids showed higher photocatalytic activity in organic dye degradation than their separate components. Notably however, another study reported reduced photocatalytic dye degradation activity from hybridizing $Bi_2O_2CO_3$ to $TiO_2$ compared to neat $TiO_2$.[61] In either case, no work has as of yet hybridized atomically-thin 2D $Bi_2O_2CO_3$ with $TiO_2$ or studied $Bi_2O_2CO_3$/$TiO_2$ hybrids for photocatalytic HER, as we do here.

For photocatalytic HER measurements via water splitting a sacrificial agent, which is often an alcohol, is commonly used as a hole scavenger.[62–65] One of the most commonly used systems in photocatalytic HER is a 1:1 $H_2O$:MeOH mixture.[62,63] Therefore for the 2D $Bi_2O_2CO_3$ nanoflakes/-sheets exfoliated in MeOH and $H_2O$, respectively, the respective other solvent was added towards obtaining 2D $Bi_2O_2CO_3$ nanoflakes/-sheets in 1:1 $H_2O$:MeOH mixture. This consistency of sacrificial agent concentration ensures direct quantitative comparability of HER results for MeOH and $H_2O$ preparations (within limits of the estimated maximum solid content concentrations of maximum of ~50 mg/l). To elucidate possible roles of the 2D $Bi_2O_2CO_3$ in photocatalysis, HER measurements were undertaken for i. neat 2D $Bi_2O_2CO_3$ nanoflakes/-sheet and ii. 2D $Bi_2O_2CO_3$ nanoflakes/-sheets hybridized with $TiO_2$ nanoparticles (Degussa P25, particles size ~25 nm). $TiO_2$ nanoparticles are archetypical photocatalysts, but are comparatively inactive towards HER, thus commonly requiring a HER co-catalyst such as expensive Pt.[52,53,62] Thereby combination of 2D $Bi_2O_2CO_3$/$TiO_2$ to hybrids allows to assess if 2D $Bi_2O_2CO_3$ can act as such HER co-catalyst. Notably most of the few prior reports on HER with $Bi_2O_2CO_3$ hybrids had Pt as additional co-catalyst present in the hybrids,[34,51] thus masking intrinsic $Bi_2O_2CO_3$ HER (co-)catalytic performance.

Fig. 3a shows that for neat, small 2D $Bi_2O_2CO_3$ nanoflakes (MeOH ICE) we observe no appreciable photocatalytic HER response. This suggests that under our conditions, our

2D Bi$_2$O$_2$CO$_3$ nanoflakes are not active as a monolithic photocatalyst for HER. When however hybridizing our 2D Bi$_2$O$_2$CO$_3$ nanoflakes (MeOH ICE) with TiO$_2$ nanoparticles, we notably get a significant HER response upon UV illumination (Fig. 3a). Most importantly, this HER response is significantly larger than the response from neat TiO$_2$ alone (Fig. 3a). This suggests that our small 2D Bi$_2$O$_2$CO$_3$ nanoflakes here act as useful co-catalysts for photocatalytic HER. TEM of the MeOH ICE 2D Bi$_2$O$_2$CO$_3$ nanoflake/TiO$_2$ nanoparticle hybrids after HER (Fig. 3b) indicates that TiO$_2$ nanoparticles have well adhered to the 2D Bi$_2$O$_2$CO$_3$ nanoflakes (MeOH ICE). Notably, the small 2D MeOH ICE nanoflakes are also well size matched to the TiO$_2$ nanoparticles.

Fig. 3c shows similar HER photocatalysis data for the large 2D Bi$_2$O$_2$CO$_3$ nanosheets (H$_2$O RT) without/with TiO$_2$. Similar to the small MeOH ICE nanoflakes, we find no HER activity for neat, large H$_2$O RT nanosheets. Again, we find however a significant synergetic effect of 2D Bi$_2$O$_2$CO$_3$ nanosheet (H$_2$O RT)/TiO$_2$ hybridisation towards a HER photocatalytic response clearly above the performance of neat TiO$_2$. Post-HER TEM likewise indicates good interfacing of the large 2D Bi$_2$O$_2$CO$_3$ nanosheets with TiO$_2$ nanoparticles (Fig. 3d).

Our work thereby introduces 2D Bi$_2$O$_2$CO$_3$ nanoflakes/-sheets as noble-metal-free co-catalysts with TiO$_2$ for photocatalytic HER from water splitting. Comparing the obtained specific activities from our 2D Bi$_2$O$_2$CO$_3$ nanoflakes/-sheets (with TiO$_2$) towards photocatalytic HER in our conditions (~25 μmol$_{H2}$·h$^{-1}$·mg$_{co\text{-}catalyst}^{-1}$·(W/cm$^2_{UV}$)$^{-1}$) suggests that the 2D Bi$_2$O$_2$CO$_3$ nanoflakes/-sheets facilitate the same order of magnitude of photocatalytic HER activity from UV as recently investigated non-noble co-catalysts such as Ni(O$_x$) or Cu(O$_x$) with TiO$_2$ and are more active than, e.g., Mn(O$_x$), Co(O$_x$) and Fe(O$_x$) with TiO$_2$ under similar measurement conditions.[64,65]

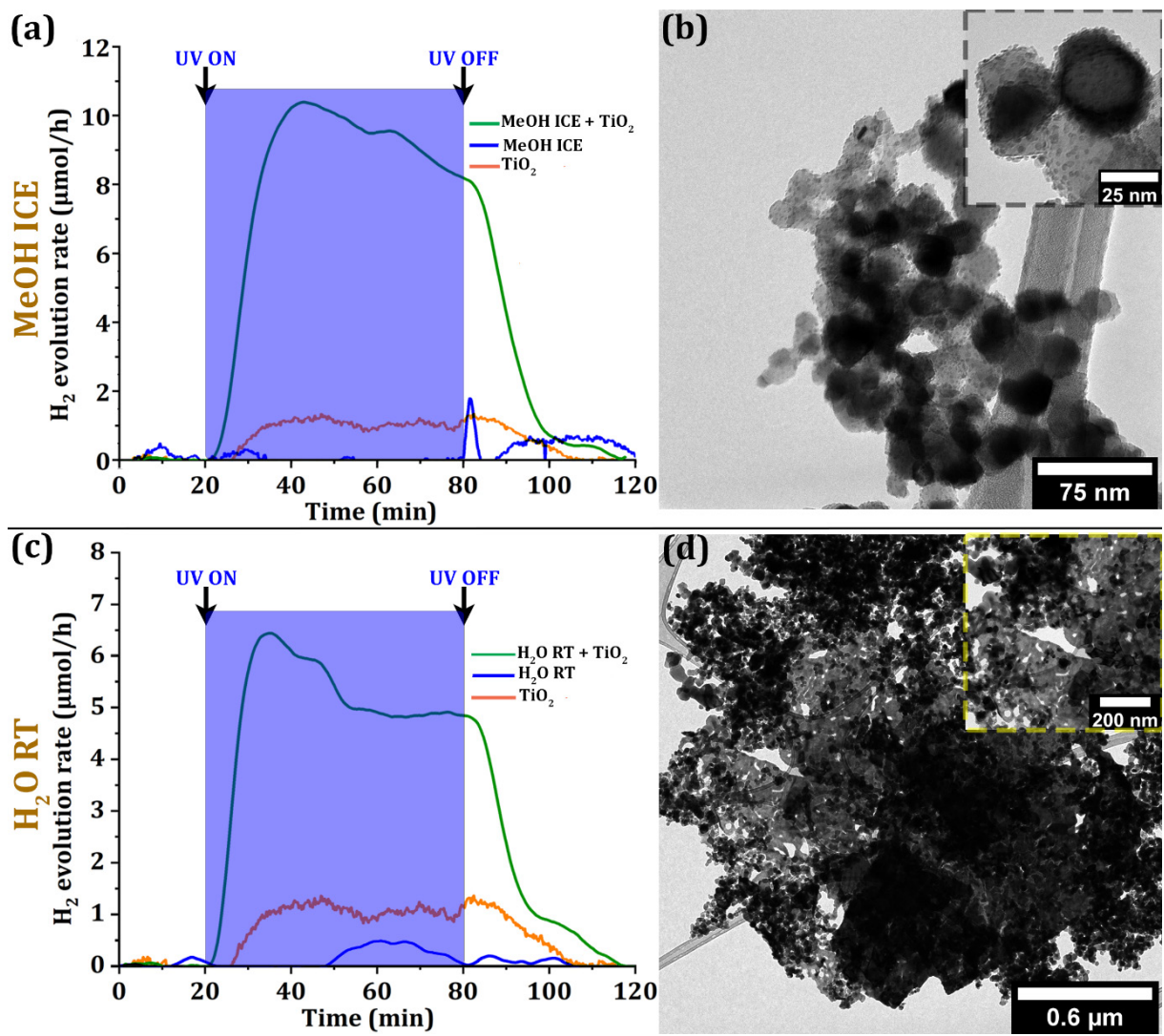

**Figure 3.** (a) Time-resolved hydrogen evolution rate during the photocatalytic HER experiments for neat MeOH ICE nanoflakes, neat TiO$_2$ nanoparticles and MeOH ICE nanoflakes hybridized with TiO$_2$ nanoparticles. (b) TEM micrograph of MeOH ICE nanoflakes hybridized with TiO$_2$ after HER measurements. (c) Time-resolved hydrogen evolution rate for neat H$_2$O RT nanosheets, neat TiO$_2$ nanoparticles (replot from (a)) and H$_2$O RT nanosheets hybridized with TiO$_2$ nanoparticles. (d) TEM micrograph of H$_2$O RT nanosheets hybridized with TiO$_2$ after HER measurements. Start and stop of UV illumination in (a,c) is indicated.

# Conclusion

In summary, we present a fabrication route which allows template-free synthesis of 2D $Bi_2O_2CO_3$ with good control over lateral 2D size, ranging from unusually large µm nanosheets to few tens of nm small nanoflakes just based on the suspension medium used. This critically adds to the toolbox for 2D pnictogen compound fabrication with emerging applications in (photo-)catalysis, electronics and energy. As one application example, we test the newly introduced 2D $Bi_2O_2CO_3$ nanosheets/-flakes in photocatalysis for sustainable solar fuel generation. In particular, our work demonstrates that the 2D $Bi_2O_2CO_3$ nanosheets/-flakes can be an advantageous co-catalyst when hybridized with $TiO_2$ for the important HER in photocatalytic water splitting.

**Acknowledgements**
B.C.B., T.G. and D.E. acknowledge partial support to this work from the Austrian Research Promotion Agency (FFG) and the Austrian Federal Ministry for Climate Action, Environment, Energy, Mobility, Innovation and Technology (BMK) under the "Production of the Future" project 883865-SolarSulfide2D. S.M.S. acknowledges partial support to this work from the European Commission (ErBeStA, No. 800942) and the Austrian Academy of Sciences ÖAW (Quantum Light, No. 1847108). We also acknowledge the use of facilities at the University Service Centre for Transmission Electron Microscopy (USTEM), Analytical Instrumentation Centre (AIC) and X-Ray Centre (XRC), TU Wien, Austria for parts of this work.

**Author contributions**
T.G., D.E. and B.C.B. planned the experiments. T.G. and N.R. carried out synthesis. T.G. performed TEM, SAED, SEM and EDX characterization. N.R. carried out the photocatalysis experiments together with T.G.. Authors M.S. and A.F. performed XPS, A.G. and S.M.S. PL and M.W. and G.F. AFM measurements. J.R. and S.N. contributed to XRD and D.A. to solvent characterization with T.G.. Author T.G. analysed all data with input from B.C.B.. Authors T.G. and B.C.B. wrote the manuscript with inputs from all authors. B.C.B. and D.E. supervised the work.

**Declaration of Competing Interest**
The authors declare no known competing interests.

**Data Availability**
The authors declare that the data supporting the findings of this study are available within the paper and its supplementary information files.

Supporting Information to:

# Sonochemical synthesis of large two-dimensional $Bi_2O_2CO_3$ nanosheets for hydrogen evolution in photocatalytic water splitting

Tushar Gupta,[1] Nicole Rosza,[1] Markus Sauer,[2] Alexander Goetz,[3] Maximilian Winzely,[4] Jakob Rath,[1] Shaghayegh Naghdi,[1] Dogukan H. Apaydin,[1] Gernot Friedbacher,[4] Annette Foelske,[2] Sarah M. Skoff,[3] Bernhard C. Bayer,[1,*] Dominik Eder[1,*]

[1]Institute of Materials Chemistry, Technische Universität Wien (TU Wien), Getreidemarkt 9/BC, Vienna, A-1060, Austria

[2]Analytical Instrumentation Center, Technische Universität Wien (TU Wien), Lehargasse 6/BL, Vienna, A-1060, Austria

[3]Atominstitut, Technische Universität Wien (TU Wien), Stadionallee 2, Vienna, A-1020, Austria

[4]Institute of Chemical Technologies and Analytics, Technische Universität Wien (TU Wien), Getreidemarkt 9/BB, Vienna, A-1060, Austria

*Corresponding authors: bernhard.bayer-skoff@tuwien.ac.at (B. C. Bayer), dominik.eder@tuwien.ac.at (D. Eder)

**Materials and Methods**

**Sonication and centrifugation of bulk Bi particles to 2D $Bi_2O_2CO_3$.** The schematic of the overall synthesis process is described in Supporting Fig. S1. The starting bulk Bi precursor was Bi powder (Goodfellow, 99.999% purity, maximum particle size of 150 µm, SEM in Supporting Fig. S2a). XRD (Supporting Fig. S2b) confirms the precursor's phase purity as phase-pure β-Bi/A7 with no oxides.

**Solvents (incl. discussion of purity).** Five different solvents were tested for their efficacy towards sonication of bismuth powder, namely: Water ($H_2O$), methanol (MeOH), isopropanol (IPA), ethanol (EtOH) and (motivated by prior literature[1]) $H_2O$:IPA (4:1) mixture. Solvents were of technical grade, stored in contact to ambient air and were used without particular purging to remove dissolved gases before synthesis. $H_2O$ was nominally deionized, however resistivity measurements on the order of only ~0.4 MΩ·cm indicate significant concentrations of residual ions to be present in the water incl. carbonate ions from $CO_2$ dissolution from ambient. Consistent with high content of residual ions in the water, we also find significant traces of Ca and Na from the $H_2O$ preparation in the XPS survey scan in Supporting Fig. S12, which we also ascribe to residual ions in the $H_2O$.

**Synthesis procedure.** The initial starting concentration of Bi was chosen to be 4 mg/ml in all solvents, where 60 mg of Bi powder was dispersed in 15 mL of solvent. Sonication employed a commercial ultra-sonic bath system (SONOREX DIGITEC DT255 H-RC ultrasonic bath manufactured by BANDELIN Electronic GmbH & Co. KG) with an operating ultrasonic frequency of 35 kHz and ultrasonic nominal power of 160 W. Sonication was carried out for 15 h. A handheld digital thermometer (Model HH11B series, single input, Type-K (Chromium-Alum) thermocouple with an accuracy of ±0.1% reading manufactured by OMEGA Engineering GmbH) was employed for the regular monitoring of the temperature of the sonication bath water to carry out the sonication process strictly within a preselected temperature range. The "room temperature" (RT) sonication was carried out at 22 °C to 26 °C, whereas during the "ice sonication" (ICE) procedure the temperature was maintained between 0.1 °C to 2.2 °C. Bi powder was weighed and put into the solvents in vials in ambient air. Then sonication was done by immersion of the air-tight capped vials with the sample solutions in the temperature-controlled sonication water bath. After the sonication, the supernatant in the vials was immediately collected and transferred to the centrifuge tubes without disturbing the sediment. From the initial 15 ml volume of the solution used for the sonication, an approximate volume of 10 ml was safely collected for the next step of centrifugation. Centrifugation was carried out in a commercial laboratory centrifuge (Model 1-6P, manufactured by SIGMA Laborzentrifugen GmbH) at the maximum speed of 5630 rpm for 20 min. After the centrifugation, again the supernatant was immediately separated without disturbing the sediment and transferred to a separate tube. From the initial 10 ml volume of the sonicated sample used for centrifugation, approximately 7.5 ml volume of supernatant was safely recovered after centrifugation. It was this volume of the supernatant (after sonication and centrifugation) that was further utilized for characterization and photocatalytic testing.

**Materials characterization.** The sonicated and centrifuged materials were characterized after drop-casting onto $SiO_2$ (90 nm) coated single-crystalline Si(100) wafers (SEM, AFM, XRD, XPS, PL) and lacey carbon TEM grids (TEM, SAED, EDX). TEM and SEM were employed for studying morphology, size distribution, thickness, crystallinity and identification of the phase (by FT and SAED patterns). The TEM characterizations were also performed on the samples post photocatalytic HER to study the hybridization of the 2D $Bi_2O_2CO_3$ with the $TiO_2$ nanoparticles. TEM studies incl. BF imaging, SAED, EDX were performed on a FEI TECNAI F20 operated at 60 kV electron acceleration voltage, sample vacuum of $\sim10^{-7}$ mbar and beam current densities of $\sim2\times10^1$ e–$Å^{-2}s^{-1}$ for imaging and SAED. The SEM measurements employed a FEI Quanta 250 FEG SEM. XRD employed a PANalytical X'Pert Pro multi-purpose diffractometer (MPD, CuKα, Bragg-Brentano geometry). Bi powder samples were deposited for XRD onto off-angle cut Si single crystal sample holders, 2D $Bi_2O_2CO_3$ drop cast onto $SiO_2$/Si wafers was measured in normal Bragg Brentano geometry. EDX employed a FEI Quanta 250 FEG-SEM via EDAX SDD Octane Elite 55 detector with $Si_3N_4$ window at 20 kV. The nanoflakes/-sheets were further subjected to AFM for determining their thickness. AFM studies employed a Brucker Nanoscope 8 multimode scanning probe microscope in tapping mode and the analysis was done using Gwyddion software.[2] XPS was conducted with a custom-built SPECS XPS-spectrometer equipped with a monochromatic Al-Kα X-ray source (µFocus 350, spot size: 400 µm, power: 100 W) and a hemispherical WAL-150 analyser (Acceptance angle 60°). The overview and detailed spectra were recorded with pass-energies of 100 eV and 30 eV, and energy resolutions of 1.0 eV and 0.1 eV, respectively. These were analysed using transmission corrections (as per manufacturer's specifications), Tougaard backgrounds,[3] and sensitivity parameters after Scofield[4,5] within CASA XPS software, and charge correction was applied so the binding energy value of adventitious carbon from surface contamination was shifted to 285 eV. Deconvolution of XPS signals was carried out using symmetric Gaussian-Lorentzian peaks (GL(30)) and Levenberg-Marquardt least-square peak fitting. PL measurements were performed using a custom-built confocal microscope with a 50 x Mitutoyo objective (NA= 0.55) and an Andor Technology Shamrock SR-303i spectrometer. The excitation laser was a 532 nm extended cavity diode laser with typical powers of 10 µW at the target. Gravimetric solid content concentration was attempted with a commercial lab balance by solvent evaporation on a hot plate, however we found concentrations were too low to allow reliable measurements. We thus estimated the maximum concentration of our nanosheets/-flakes to the minimum sensitivity via the minimum sensitivity of the lab balance to ~50 mg/l (but note that actual concentration could be below this value).

Phase analysis of TEM data (FTs of lattice resolution BF-TEM and SAED) employed primarily FT/SAED pattern simulation using Highscore Plus/PDF-4+ software[6] (ICDD PDF-4+ 2021 RDB: Software version: 4.21.0.2. Database version: 4.2103.) for manual matching of measured and simulated FT/SAED patterns. Structure visualization was done by Vesta[7] software. In particular the following structural Powder Diffraction File (PDF) database entry was found to best fit our measured FTs and SAEDs (PDF-4+ code): orthorhombic 04-009-8533(previous PDF code: 01-084-1752).[8] We comment on alternative descriptions via other PDF files for $Bi_2O_2CO_3$ below, incl. other orthorhombic and (often also used) tetragonal PDF files. With regard to other possible phase matches, we checked 38 Bi and 101 $Bi_xO_y$ entries from the ICDD PDF-4+ database which consistently gave worse matches to experimental data. A whole section below is devoted to discussion of the in-depth structural phase analysis.

**Photocatalytic HER screening.** For photocatalytic HER measurements in a suspension-type reactor, first appropriate reaction solutions were prepared which included as sacrificial agent MeOH in 1:1 $H_2O$:MeOH mixture.[9–12] Thus for $H_2O$ and MeOH nanoflake/-sheet preparations, the respective other solvent was added to obtain nanoflakes/-sheets in 1:1 $H_2O$:MeOH. This consistency of sacrificial agent concentration ensures direct quantitative comparability of HER results for $H_2O$ and MeOH samples (within limits of only estimated maximum concentration of ~50 mg/l, see above). HER measurements were undertaken with neat 2D $Bi_2O_2CO_3$ and 2D $Bi_2O_2CO_3$ hybridized with $TiO_2$ nanoparticles (P25 from Degussa), which are well known HER photocatalysts (7.5 mg $TiO_2$ nanoparticles per 30 ml HER solution).[9]

For photocatalytic HER measurements suspensions were sonicated for 3 min after mixing, after which solutions were transferred to suspension-type HER reactors[9–12] (30ml) under constant stirring. In the reactor the HER solution was purged with Ar flow to eliminate the dissolved oxygen and also to ensure the proper mixing. Afterwards, the reactor was closed. During light irradiation, there was constant Ar flow (30 ml / min) controlled by a mass flow controller (Q-flow 140 series, MCC-Instruments) carrying all the gases from the reactor to the gas detector. Hydrogen generation was detected online and in-stream using an Emerson gas analyzer equipped with a thermal conductivity detector. The light source used was a 0.55 W output power LED (SOLIS-365C by THORLABS, Inc.), which provides UV-irradiation with a nominal wavelength of 365 nm. Estimated illuminated area on the reactor is ~1 cm$^2$.

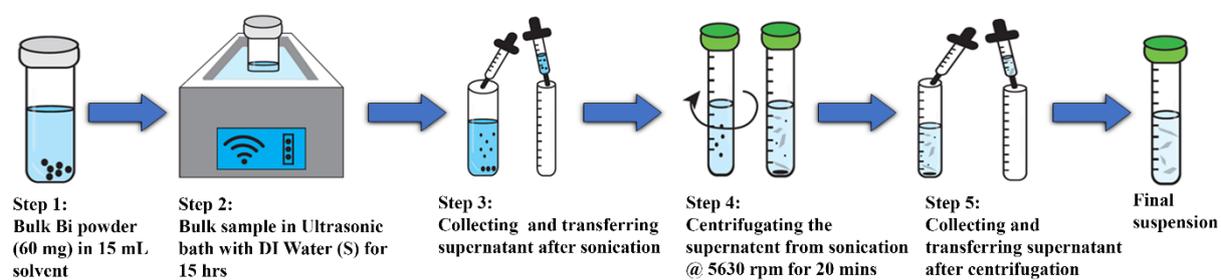

**Supporting Figure S1.** Schematic diagram of the synthesis methodology incl. ultrasonication and centrifugation steps.

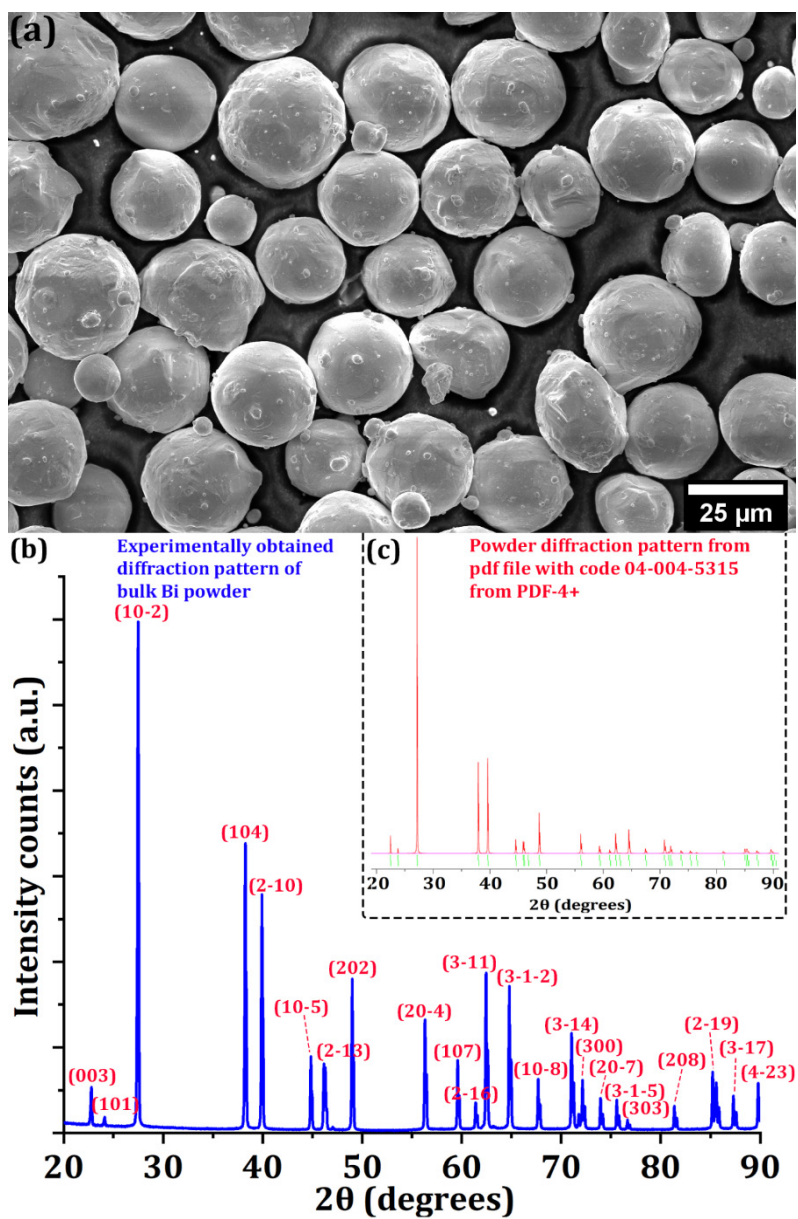

**Supporting Figure S2.** (a) SEM image of initial bulk Bi powder. (b) shows the experimentally obtained diffraction pattern of the bulk Bi powder (c) shows the simulated diffraction pattern from the PDF file 04-004-5315 corresponding to rhombohedral β-Bi. The peaks in the obtained diffraction pattern in (b) have been (hkl)-indexed to the simulated pattern in (c). All the peaks obtained in the diffractogram were successfully indexed to β-Bi. This reaffirms the phase purity of bulk Bi powder and verifies the absence of any Bi-oxides in the initial precursor before our sonochemical synthesis.

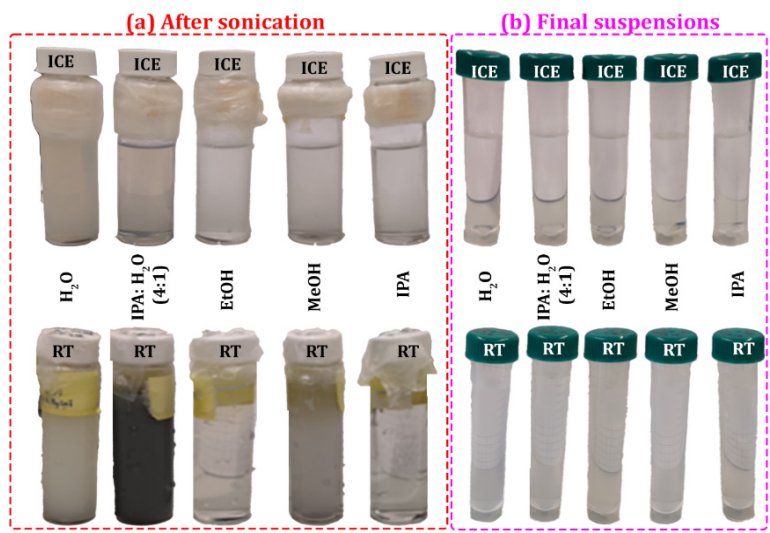

**Supporting Figure S3.** Photographs showing the optical appearance of the suspensions of 60 mg of Bi powder in 15 mL of the solvents after 15 h of sonication (a) and of the supernatant collected after subsequent centrifugation (b).

**Supporting characterization data**

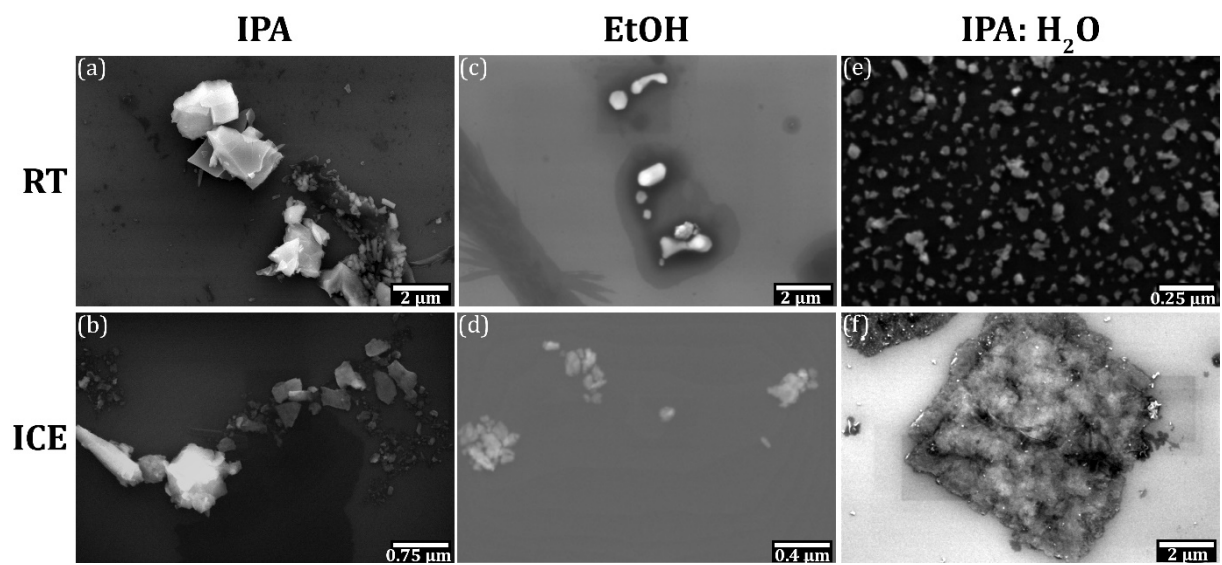

**Supporting Figure S4.** SEM images of obtained nanoflakes in IPA (a,b); EtOH (c,d); IPA: H$_2$O (4:1) (e,f) at RT (a,c,e) and ICE conditions (b,d,f), respectively.

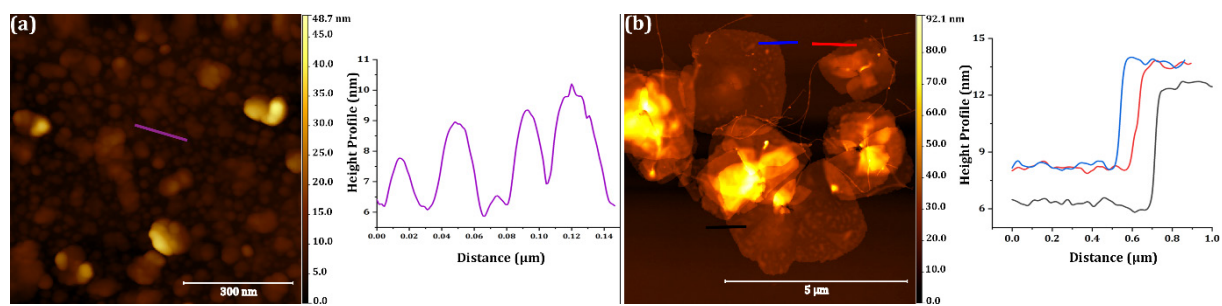

**Supporting Figure S5.** AFM images of MeOH RT (a) and H$_2$O RT (b). Line profiles over the flakes are plotted alongside to measure flake thickness.

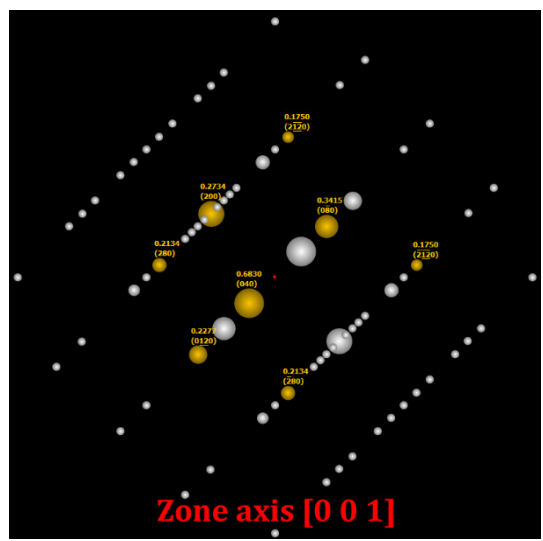

**Supporting Figure S6.** Side view FT/SAED simulation for orthorhombic $Bi_2O_2CO_3$ (PDF file orthorhombic 04-009-8533[8]). Note that FT/SAED simulation is not rotation corrected with respect to side-view, i.e. "basal" (010) plane family parallel to substrate appear inclined at ~45 ° in this simulation.

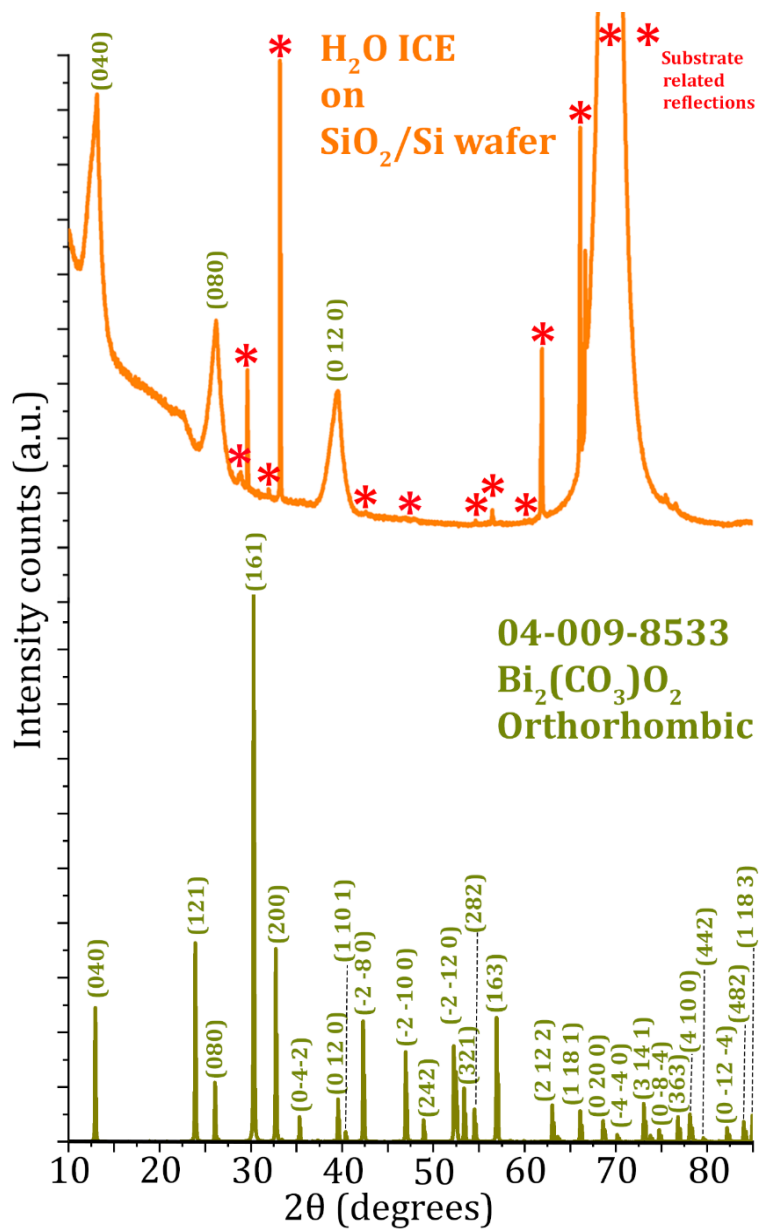

**Supporting Figure S7.** XRD pattern (Bragg-Brentano geometry) of H$_2$O ICE drop cast onto SiO$_2$/Si substrates. XRD peaks related to the the single-crystalline Si(100) wafer substrate are marked with red asteriks (*). Three reflections (~13°, ~26° and ~40°) are related to the H$_2$O ICE nanosheets and are consistently indexed to orthorhombic Bi$_2$O$_2$CO$_3$ (PDF: 04-009-8533[8]) (040), (080) and (0120) reflections (as labelled), respectively. This is for XRD in Bragg-Brentano geometry exactly the pattern one would expect from 2D Bi$_2$O$_2$CO$_3$ with (010) texture, as indicated from TEM data in the main text. See below (Supporting Fig. S20) for an extended description of XRD phase assignment also with respect to other Bi$_2$O$_2$CO$_3$ PDF file entries (orthorhombic and tetragonal) and other candidate Bi and Bi$_2$O$_3$ phases (which all consistently give worse matches than Bi$_2$O$_2$CO$_3$).

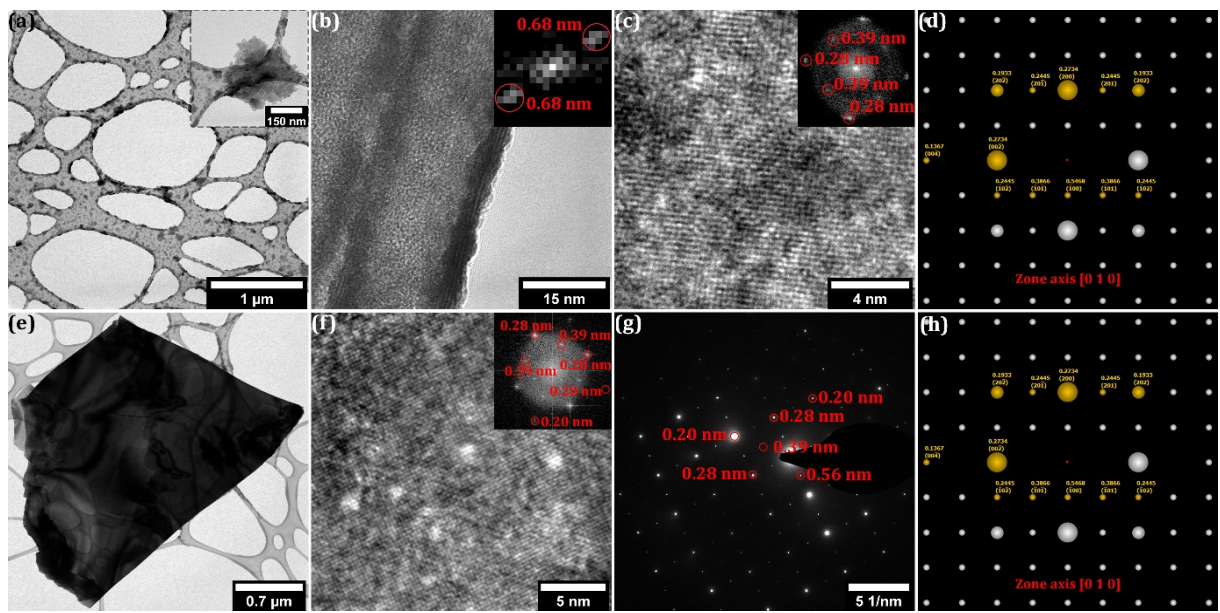

**Supporting Figure S8.** (a-d) MeOH RT studied by overview BF-TEM (a), lattice resolution side-view (b, FT inset) and lattice resolution top-view (c, FT inset) of flake. (d) shows a simulated SAED/FT pattern for orthorhombic $Bi_2O_2CO_3$ along [010] zone axis with salient reflections matches to experimental data in (c) highlighted. (e-h) $H_2O$ RT studied by overview (e) and lattice resolution (f, FT in inset, top-view of flake) BF-TEM and top-view SAED (g). (h) shows a simulated SAED/FT pattern for $Bi_2O_2CO_3$ along [010] zone axis with salient reflections matches to experimental data in (f,g) highlighted.

**EDX mapping.** To verify homogeneous presence of Bi, O and C in the nanosheets, EDX mapping of $H_2O$ ICE nanosheets on amorphous carbon TEM membranes was undertaken. Elemental mapping clearly shows the homogeneous presence of Bi, O and C in the nanosheets, corroborating that the nanosheets are 2D $Bi_2O_2CO_3$ (and not metallic 2D bismuthene).

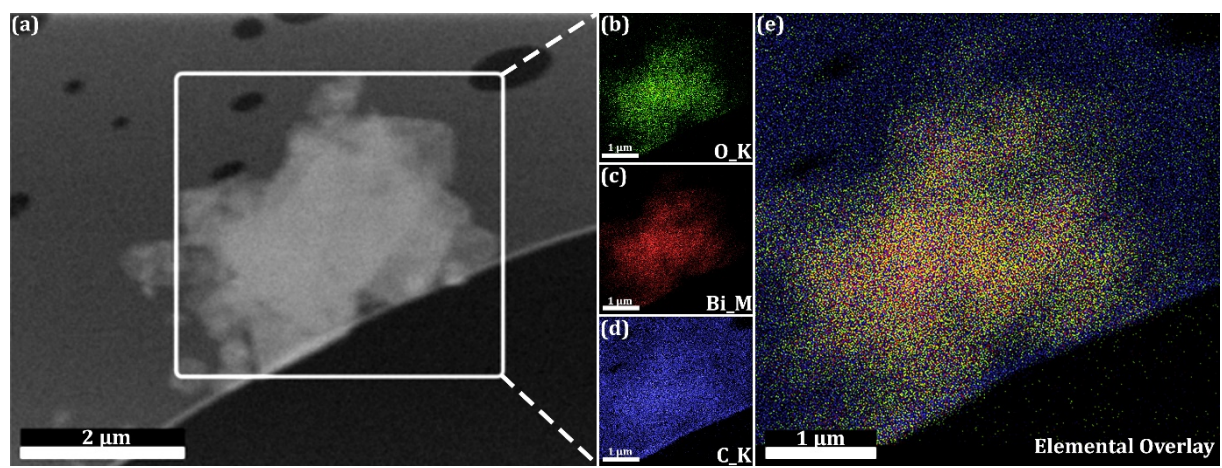

**Supporting Figure S9.** Qualitative analysis of Bi $H_2O$ ICE nanosheets via EDX mapping. (a) shows a SEM micrograph of the region of interest in the white square in which EDX mapping was done. (b), (c) and (d) show the laterally resolved EDX maps for oxygen, bismuth and carbon respectively. A corresponding, elemental overlay composite image of the three elemental maps is presented in (e).

**XPS.** XPS measurements were done on $H_2O$ ICE nanosheets on $SiO_2$ (90 nm)/Si wafer. The Bi4f signal has only a single component at a binding energy (BE) of 159.2 eV ($Bi4f_{7/2}$) consistent with $Bi^{3+}$ (as in, e.g., $Bi_2O_2CO_3$, Supporting Fig. S10a).[13] The Bi4f signal in particular excludes the presence of metallic $Bi^0$ which would be expected at a BE of 156.9 eV.[14,15] Components characteristic of oxycarbonates are correspondingly found in the O1s signal (Supporting Fig. S10b) at a BE of 529.8 eV for Bi-O,[13] at 532.1 eV for C-O[13] (and/or additionally $SiO_2$ substrate).[16] An intermediate component at 530.9 eV is attributed to carbonate ions[13] and possibly C=O in either the deposit or adventitious carbon contamination. The C1s signal in Supporting Fig. S10c exhibits three components at 285.0 eV, 286.2 eV and 288.6 eV that can be attributed to C-C/C-H, C-O and oxycarbonates respectively, the latter possibly overlapping with the signal of C=O.[13] In particular, the 285.0 eV peak is also consistent with adventitious carbon adsorption from sample storage in ambient air. Survey XPS spectra (Supporting Fig. S11) detect only Bi, O and C plus Si from the substrate as well as traces of Na and Ca, which we link to residual ions in the $H_2O$ solvent (see also above). In summary, XPS shows signatures consistent with $Bi_2O_2CO_3$.[13]

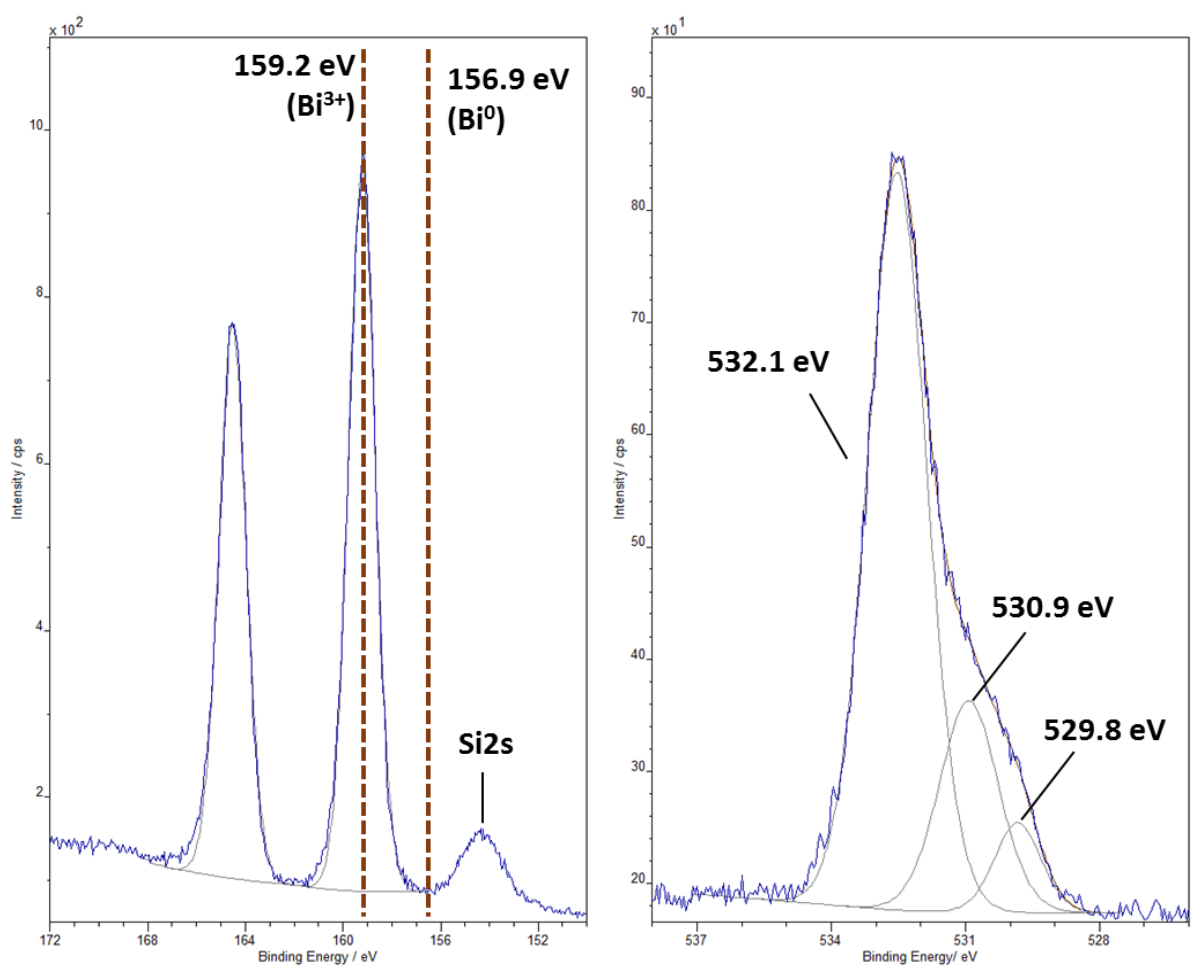

**Supporting Figure S10.** (a) Bi4f core-level spectra $H_2O$ ICE nanosheets on $SiO_2$/Si and (b) corresponding O1s spectra.

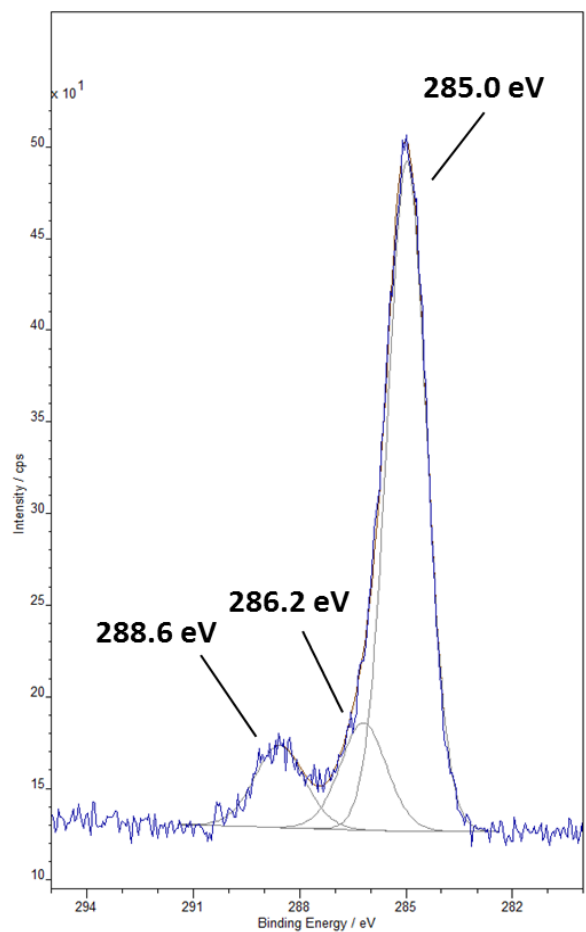

**Supporting Figure S11.** (a) C$_1$s core-level spectra H$_2$O ICE nanosheets on SiO$_2$/Si.

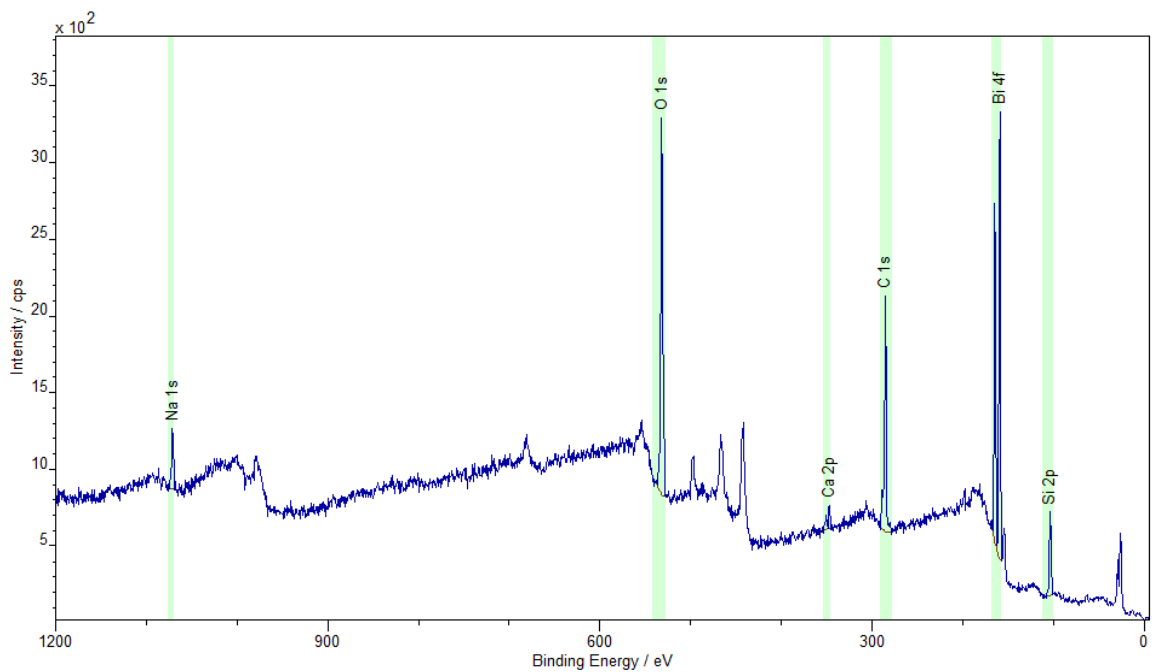

**Supporting Figure S12.** XPS survey spectrum of $H_2O$ ICE nanosheets on $SiO_2$/Si confirming the presence of Bi, O, C and Si (substrate) and additional adventitious C on the samples. Notably minor traces of Na and Ca are also consistently found from the $H_2O$ processing. We ascribe these Na and Ca signals to residual ions in the $H_2O$ solvent used, since we obtain similar measurements also for empty $H_2O$ drop-cast onto $SiO_2$/Si wafers, but not for unprocessed $SiO_2$/Si wafers (not shown). The finding of residual ions in the $H_2O$ also correlates well with the resistivity measurements of the water discussed above.

**PL.** PL mapping (from 532 nm excitation) was used to study optoelectronic properties of the $H_2O$ ICE nanosheets on $SiO_2$/Si. Supporting Fig. S13 (left) shows a laterally resolved PL emission intensity map (excitation beam of 532 nm filtered by 550 nm longpass filter), which indicates the nanosheets to have a PL response well above the "dark" $SiO_2$/Si substrate. Supporting Fig. S13 (right) shows a PL emission spectrum from the flake circled blue on the map on the left, indicating that the PL emission is predominantly from emission at ~550 nm (confirmed for several flakes, spectrum has been background corrected for $SiO_2$/Si background). PL emission at ~550 nm (for 532 nm excitation) is consistent with prior reports of PL from $Bi_2O_2CO_3$.[17,18] Similar PL emission characteristics are also measured for MeOH ICE nanoflakes (not shown).

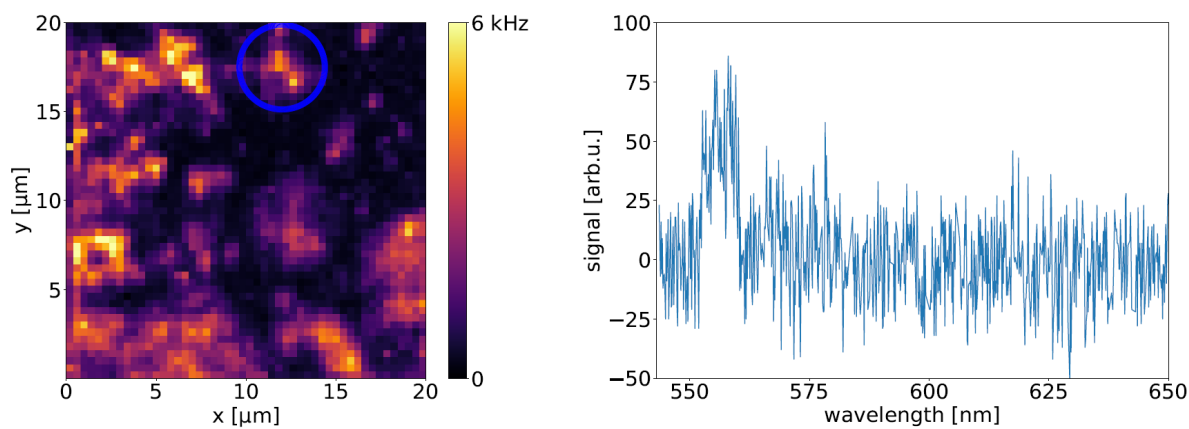

**Supporting Figure S13.** (left) latterally resolved PL emission intensity map of $H_2O$ ICE flakes on $SiO_2$/Si and (left) PL emission spectrum of a flake in the left PL emission map, background corrected over "dark" $SiO_2$/Si background.

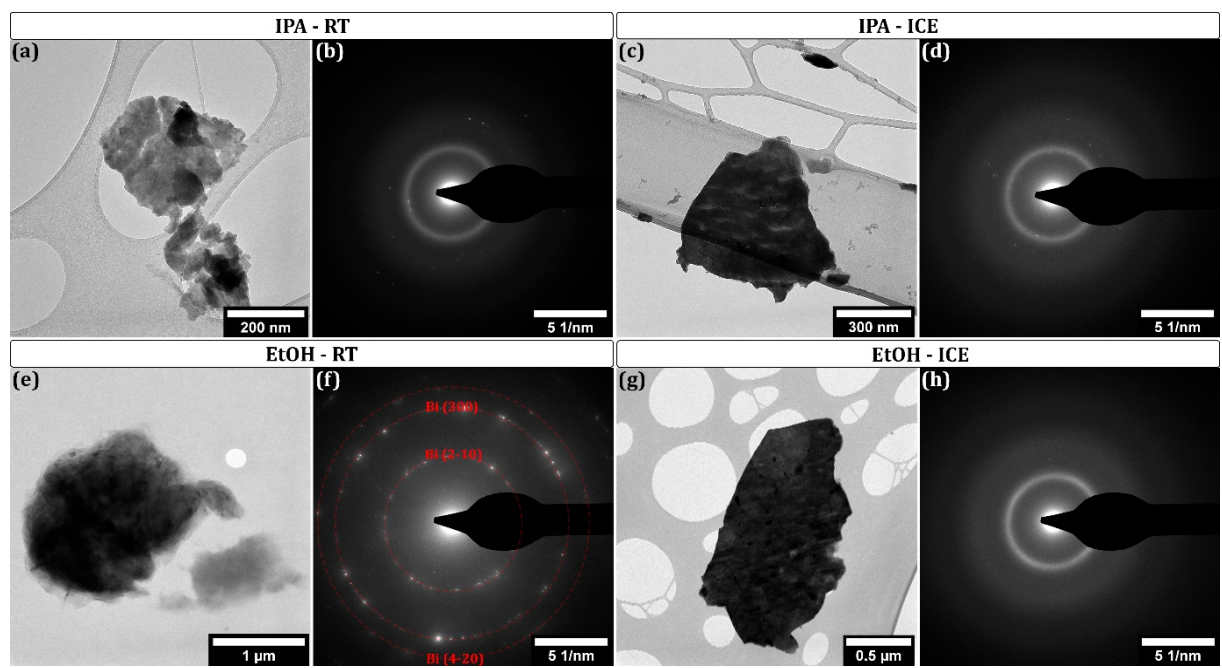

**Supporting Figure S14.** BF TEM (a) and SAED (b) of IPA RT and BF TEM (c) and SAED (d) of IPA ICE. SAED in (b) and (d) suggests amorphous structure of the nanoflakes. BF TEM (e) and SAED (f) of EtOH RT. The SAED in (f) indicates that the produced nanoflakes are of metallic β-Bi structure i.e. phase of initial β-Bi powder and no $Bi_2O_2CO_3$ phase was formed in EtOH at RT. BF TEM (g) and SAED (h) of EtOH ICE. SAED in (h) suggests amorphous structure of the nanoflakes.

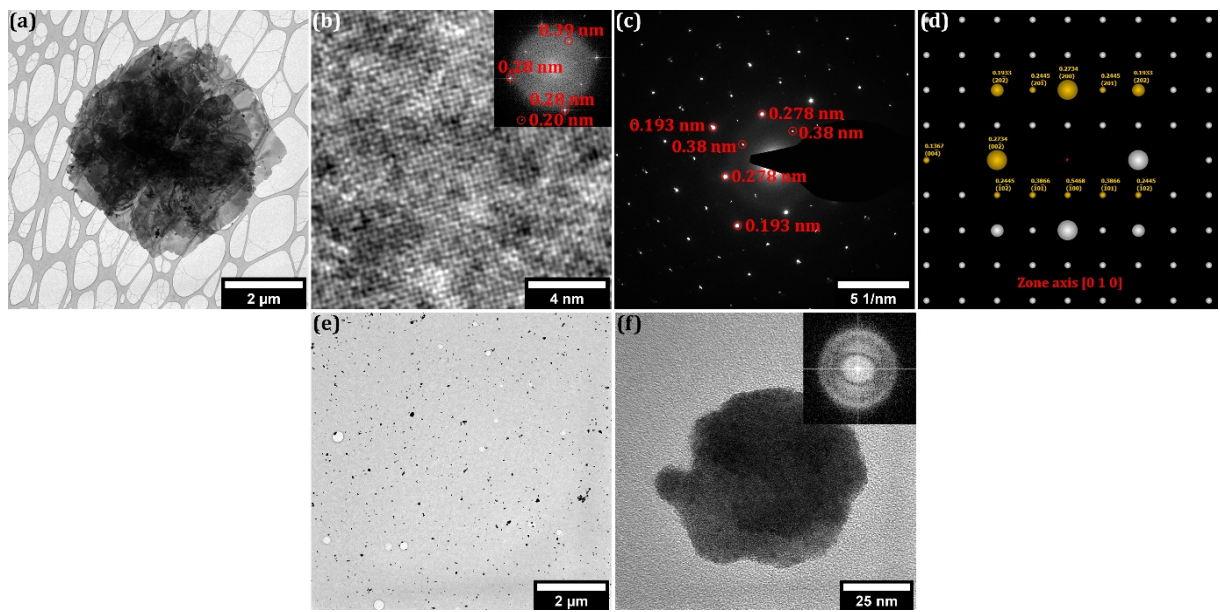

**Supporting Figure S15.** (a-d) IPA:H$_2$O ICE studied by overview (a) and lattice resolution (b, FT in inset, "top view" of flake) BF-TEM and SAED (c). (d) shows a simulated SAED/FT pattern for orthorhombic Bi$_2$O$_2$CO$_3$ along [010] zone axis with salient reflections matches to experimental data in (b,c) highlighted. (e-f) IPA:H$_2$O RT studied by overview (e) and lattice resolution (f) BF-TEM. The FT in inset shows that the obtained nanoflakes from IPA:H$_2$O RT are amorphous.

## Discussion of phase analysis of the nanoflakes/-sheets

The phase analysis of our nanoflakes/-sheets rests on lattice resolution TEM (FT) and SAED data in top-view and side-view as well as XRD data measured in Bragg-Brentano geometry.

**Top-view TEM (FT) and SAED.** We first discuss the phase identification based on the top-view TEM (FT) and SAED data (Fig. 2c,e,f and Supporting Fig. S15b,c)

Since the starting bulk material was metallic β-Bi powder without oxides, the initial screening of the phase of the synthesized nanosheets/-flakes was done with all the (semi-)metallic Bi PDF files in the PDF-4+ database. The atomic models along with the simulated diffraction patterns for the two often observed zone axes in 2D bismuth of β-Bi structure ([001] and [2-21]) are presented in Supporting Fig. S16. Both give worse matching to the d-spacings in the FT patterns of the here synthesized nanosheets/-flakes (Fig. 2c,e,f and Supporting Fig. S15b,c) compared to $Bi_2O_2CO_3$. A projection along [42-1] has also been presented additionally to underscore the differences in our results with those of the prior work by Pumera et al.[15,19] with similar but shorter sonication scheme which resulted not in 2D bismuth oxy carbonates but metallic 2D Bi. Also for this β-Bi [42-1] zone axis, our experimental data is not well matched by β-Bi. This concludes that our nanosheets/-flakes are not β-Bi. Similarly worse matches were obtained for other metallic Bi phases, excluding that our nanosheets/-flakes are any allotrope of bismuth. A total of 38 entries in the database were analyzed and no suitable match to our nanosheets/-flakes was found. (Note that for Bi atomic structure models are often described in literature with hexagonal axis (as here) but also with rhombohedral axis. Therefore, numerical (hkl) and [uvw] values need consideration of selected hexagonal or rhombohedral axis system, when comparing between reports. To avoid ambiguity the here used axis are plotted alongside the atomic models.)

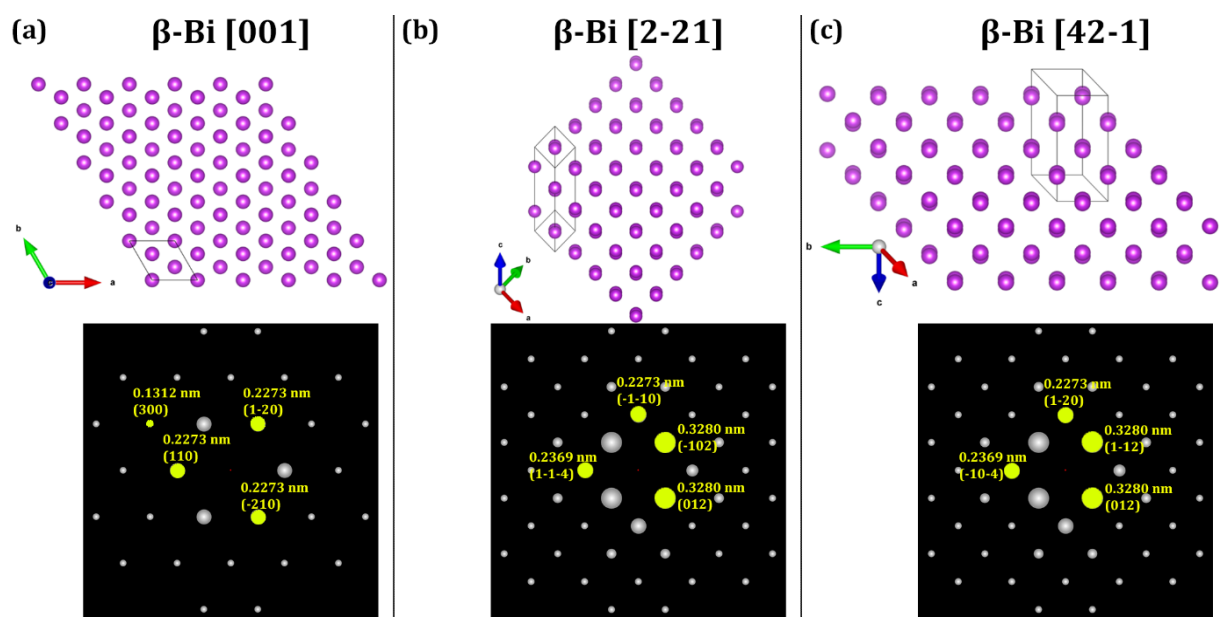

**Supporting Figure S16.** The atomic structure (top) and simulated SAED patterns (bottom) for β-Bi along (a) [001], (b) [2-21] and (c) [42-1] zone axes. The PDF file with code 04-007-5315[20] from PDF-4+ crystallographic database was used to construct the atomic and the diffraction models. (b) and (c) also represent the orientations of the β-Bi sheets obtained by Pumera et al.[15,19]

As next possible reaction product, we compared our experimental FT and SAED data to the different oxides of bismuth. A total of 102 PDF files of bismuth oxides available in PDF-4+ database with different stoichiometric ratios of Bi and O were checked. In particular, the thermodynamically most stable $Bi_2O_3$ phases were checked with respect to the atomic structure and equivalent diffraction patterns along various zone axes. Low index [001] zone axes are shown in Supporting Fig. S17 as examples of this analysis. PDF file with the atomic structure for ω-$Bi_2O_3$ could not be found. The SAED pattern for ω-$Bi_2O_3$ was simulated on the basis of the reflection intensities from the powder diffraction pattern as mentioned by the work[21] cited in the PDF file with code 00-050-1088. We note that δ-$Bi_2O_3$ along [001] zone axis also gives a very good match to our experimental top-view TEM (FT) and SAED data in Fig. 2c,e,f and Supporting Fig. S15b,c. For δ-$Bi_2O_3$ however in side view no lattice distance of 0.68 nm (as experimentally measured in Fig. 2b) is expected but rather only a much shorter distance of ~0.39 nm. This excludes δ-$Bi_2O_3$ as possible phase for our nanosheets/-flakes. Additionally, as discussed below also the measured XRD patterns of our nanosheets (Supporting Fig. S20) are inconsistent with δ-$Bi_2O_3$. In prior studies[22] particles of β-$Bi_2O_3$ were also found aligned along [110] zone axis. We therefore also cross checked the alignment of our nanosheets/-flakes with respect to β-$Bi_2O_3$ aligned along [110] zone axis as shown in Supporting Fig. S18. We note that β-$Bi_2O_3$ along [110] is also a good match for the top-view TEM and SAED data. However our XRD data in Supporting Fig. S20 is inconsistent with β-$Bi_2O_3$ (see below), thus excluding β-$Bi_2O_3$.

After having excluded Bi and Bi-oxides, we find that orthorhombic $Bi_2O_2CO_3$ with [010] texture (PDF file: 04-009-8533)[8] matches our TEM data excellently both in top-view and side-view (Supporting Fig. S19a, see also Fig. 2 and Supporting Fig. S6). Additionally, also our XRD data is perfectly matched by orthorhombic $Bi_2O_2CO_3$ with [010] texture (PDF file: 04-009-8533)[8], as shown in Supporting Fig. S20 and discussed below.

**Comment on alternative $Bi_2O_2CO_3$ unit cells.** We note that $Bi_2O_2CO_3$ unit cell has been described in various orthorhombic and tetragonal notations in the PDF file database. Notably our TEM and XRD data is also well matched by other PDF entries for $Bi_2O_2CO_3$ like the orthorhombic PDF file 04-014-4450 (Supporting Fig. S19b)[23] or the often used tetragonal description as in PDF file 00-041-1448 (Supporting Fig. S19c).[8] The reason for choosing orthorhombic $Bi_2O_2CO_3$ (PDF file: 04-009-8533) as the best description in our report is that the experimentally observed Fig. 2 and Supporting Fig. S6 ~0.38 nm reflection in FT/SAED data are best represented by the orthorhombic $Bi_2O_2CO_3$ (PDF file: 04-009-8533), while not represented in the other $Bi_2O_2CO_3$ PDF entries.

We note that between the different orthorhombic PDF files the axis orientation can be different, and there is likewise also a difference of axis orientation for the tetragonal PDF files. Therefore (hkl) and [uvw] values as used in texture description between the various PDF files vary in numerical values. We therefore for most atomic models sketches in the report always also plot the axis system used. Notably, the stated orthorhombic $Bi_2O_2CO_3$ with [010] texture in PDF file: 04-009-8533[8] is equivalent to [001] texture for orthorhombic PDF file 04-014-4450 and [001] for tetragonal PDF file 00-041-1448. As clearly shown in Supporting Fig. S19, all these descriptions (irrespective of axis system and numerical (hkl) [uvw] values) refer to $Bi_2O_2CO_3$ with the $Bi_2O_2$ sub-layers parallel to the support.

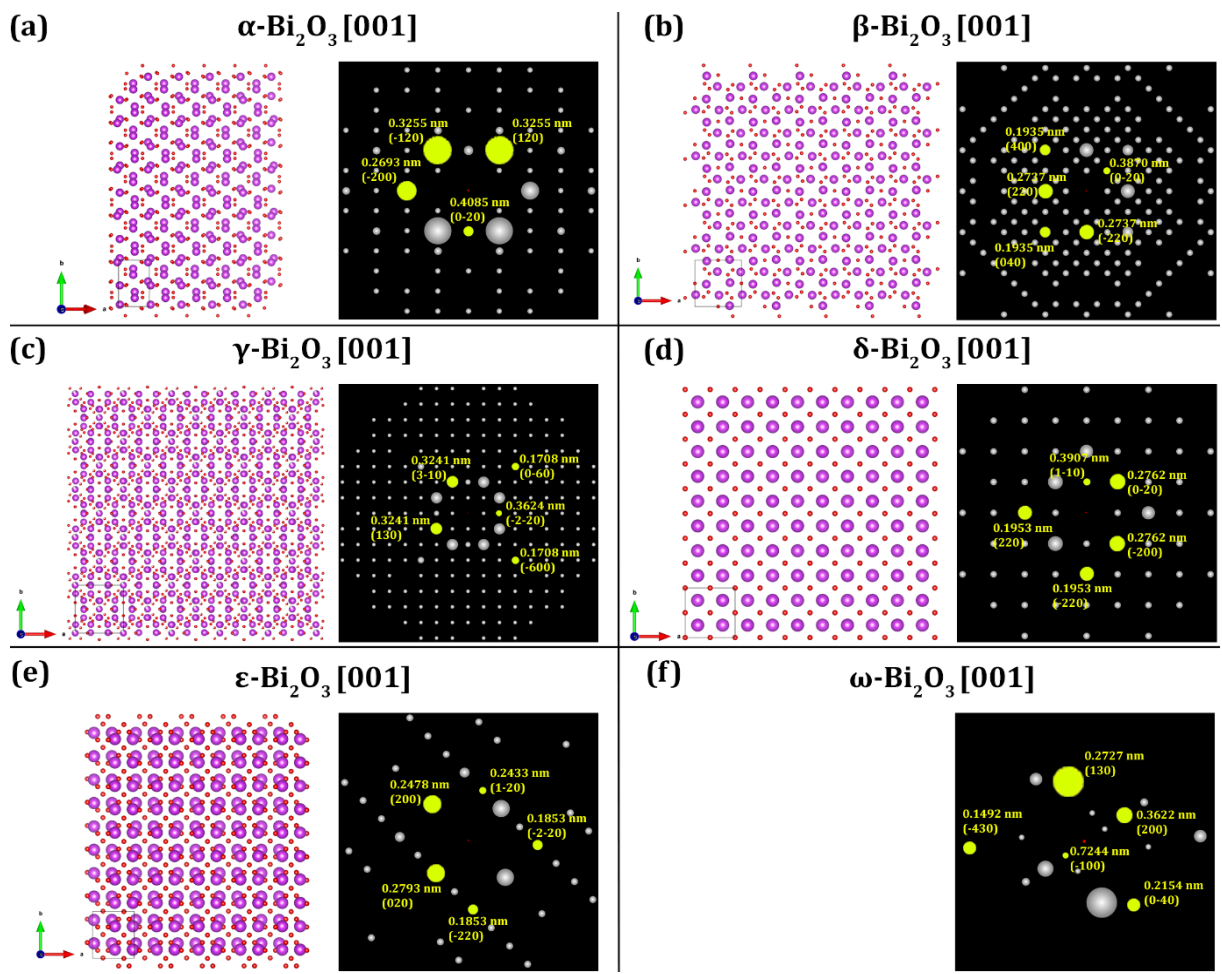

**Supporting Figure S17.** Atomic structure (left) and simulated SAED patterns (right) for 6 different allotropes (a-f) of bismuth trioxides along low index [001] zone axes as example. The atomic structure and the SAED pattern for α-$Bi_2O_3$ with monoclinic crystal system was simulated using the PDF file with code 00-041-1449[24–26] from PDF-4+ crystallographic database. For the case of β-$Bi_2O_3$ with tetragonal crystal system, PDF file with code 04-007-1443[27] from PDF-4+ database was used to reconstruct atomic and diffraction models. The structure of the γ-$Bi_2O_3$ is presented via atomic models and equivalent diffraction pattern using PDF file with code 04-007-2395[28]. PDF file with code 04-015-0028[29–31] was used to project the crystal structure of δ-$Bi_2O_3$ via atomic and diffraction models. To illustrate the orthorhombic structure of ε-$Bi_2O_3$ via corresponding atomic coordinates and diffraction profile, PDF file with code 04-013-1463[32] was utilized. For the illustration of triclinic structure of ω-$Bi_2O_3$, we could not find any entry in the database. The PDF file with code 00-050-1088[21] could only simulate the SAED profiles based on the powder diffraction data but could not provide the information about the atomic positions. [uvw] denotes the zone axis.

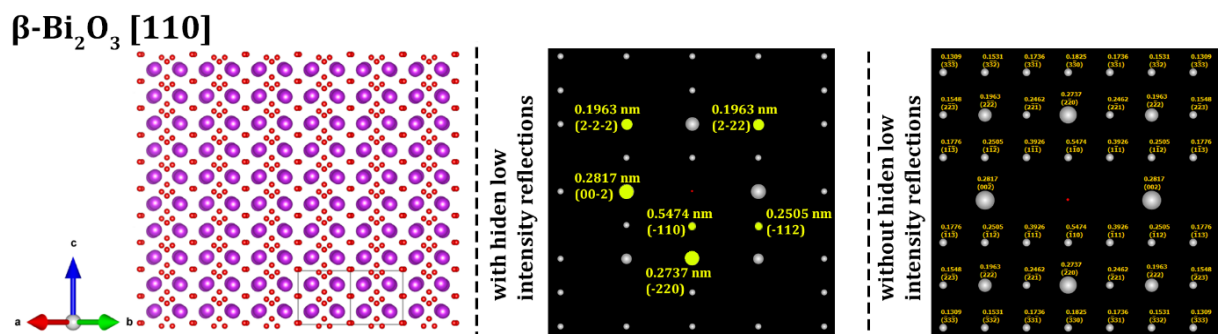

**Supporting Figure S18.** Atomic structure (left) and simulated SAED patterns (middle, right) for β-Bi$_2$O$_3$ along [110] direction constructed from PDF file with code 04-007-1443 from PDF-4+ database. The simulated FT/SAED profiles are presented in two fashions: (middle) keeping the low intensity reflections hidden and (right) with all the reflections visible to simplify the comparison for phase analysis. [uvw] denotes the zone axis.

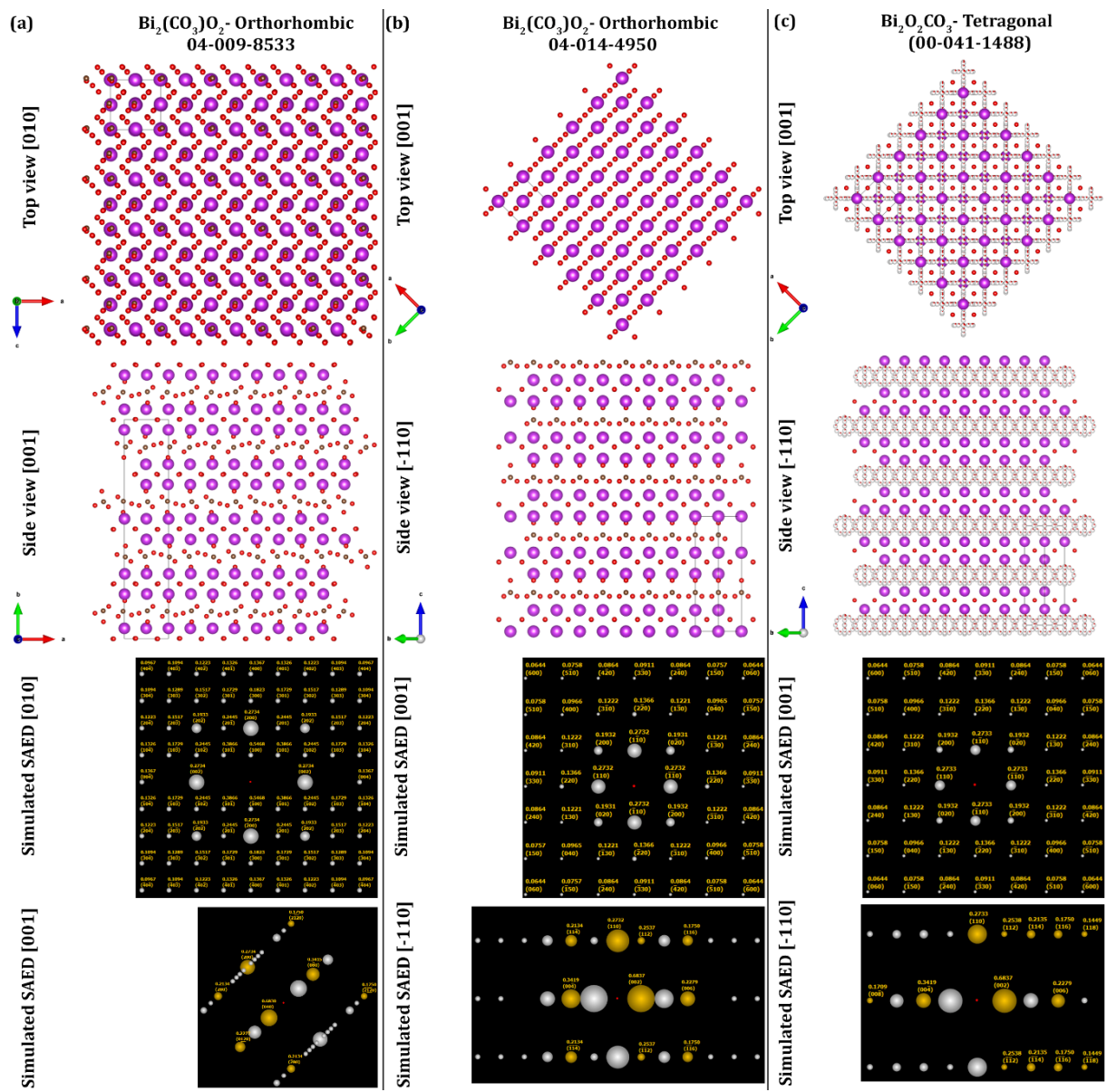

**Supporting Figure S19.** From top to bottom: Atomic structure models in top-view and side-view and corresponding simulated FT/SAED for top-view and side-view for the PDF files for $Bi_2O_2CO_3$ (a) orthorhombic 04-009-8533[8] (the majority file used in this report) and comparison (b) orthorhombic 04-014-4450[23] and (c) tetragonal 00-041-1448.[8] We emphasize that, while numerical (hkl)/[uvw] values between PDF files vary due to chosen axis system, all shown PDF entries similarly describe $Bi_2O_2CO_3$ with the $Bi_2O_2$ sub-layers parallel to the substrate. The reason for choosing orthorhombic 04-009-8533[8] as the best match in this report is the representation of ~0.38 nm reflections that we experimentally observe in FT/SAED data in Fig. 2 and Supporting Fig. S6 and which are not represented in the other PDF entries.

**XRD.** We finally present the measured XRD data compared to simulated diffraction patterns of all salient metallic Bi and Bi-oxide phases as well as $Bi_2O_2CO_3$ oxycarbonate phases (in both orthorhombic and tetragonal descriptions). As immediately apparent, the only set of XRD patterns that can account for all peaks observed in the measured XRD data (~13°, ~26° and ~40°) are the $Bi_2O_2CO_3$ patterns. This is most pronounced for the reflection at ~13 ° which is not consistent with any of the $Bi_2O_3$ or Bi phases, but only with $Bi_2O_2CO_3$. Importantly, the three observed peaks are also systematically related to each other as (040), (080) and (0120) reflections of $Bi_2O_2CO_3$ which is for XRD in Bragg Brentano geometry exactly the pattern one would expect from 2D $Bi_2O_2CO_3$ with (010) texture, that was indicated already from the TEM data. Importantly note, that none of the metallic Bi and Bi-oxide phases can account at all for the peak at 13°, thus excluding all these phases.

**Conclusion of Structural Phase Identification.** In summary, the structural assignment of our nanoflakes/-sheets to 2D $Bi_2O_2CO_3$ with (010) texture is fully consistent with TEM (FT) and SAED data in both top-view and side-view and also with Bragg-Brentano XRD data. No other candidate phases such as metallic Bi or Bi-oxides shows similar consistency in top-/side-view and XRD. This reaffirms the assignment of our nanoflakes/-sheets to 2D $Bi_2O_2CO_3$ with (010) texture.

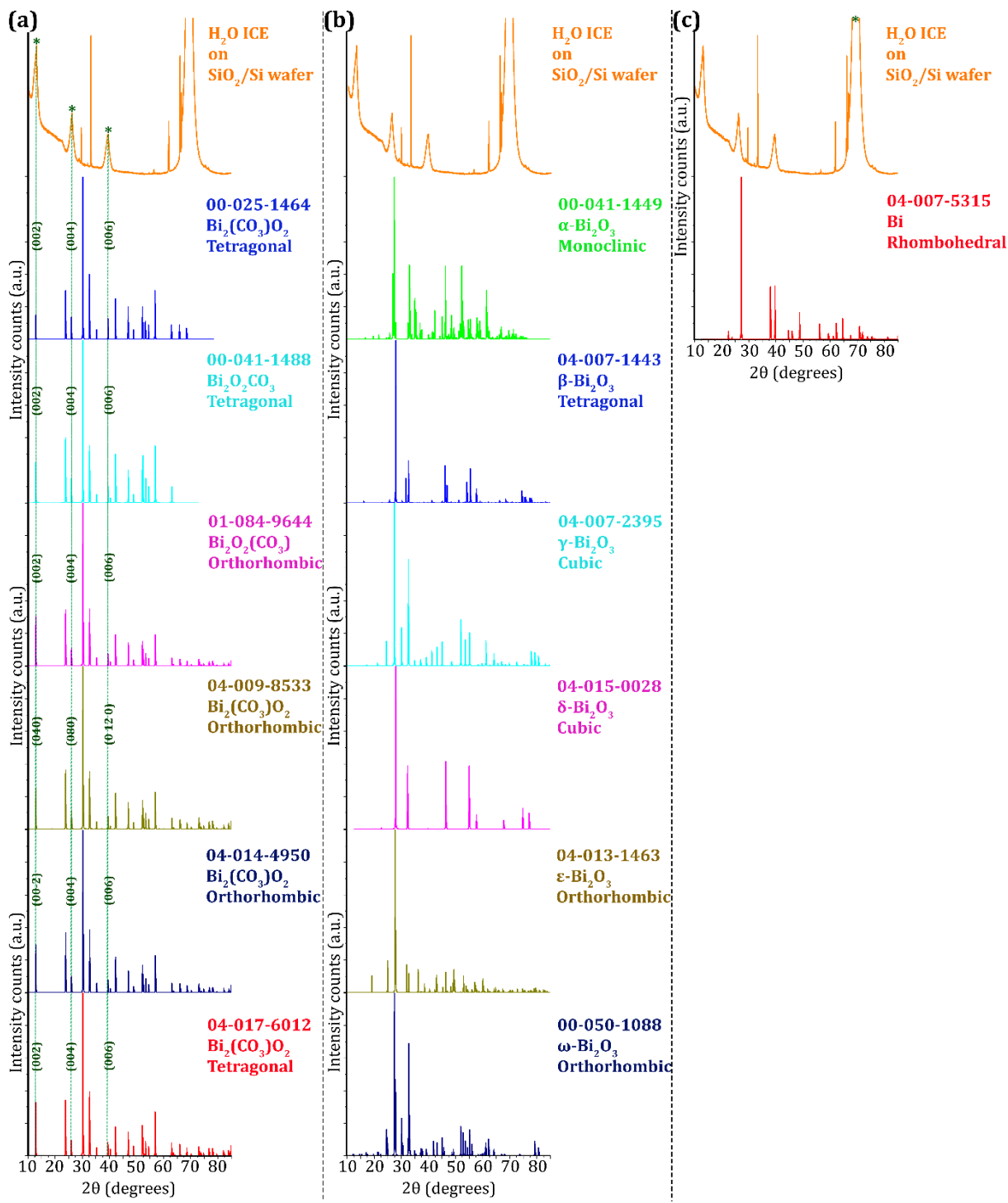

**Supporting Figure S20.** Comparison of measured XRD pattern of H₂O ICE nanosheets drop cast onto SiO₂/Si wafers (top patterns in orange) with respect to simulated powder diffraction patterns for various PDF entries (entry numbers in figure) for (a) Bi₂O₂CO₃, (b) Bi₂O₃ and (c) Bi phases.

**Supporting References**